\documentclass[journal]{IEEEtran}

\usepackage{amssymb}
\usepackage[cmex10]{amsmath}
\usepackage{amsfonts}
\usepackage{lineno}
\modulolinenumbers[5]
\usepackage{textgreek}
\usepackage{graphicx}
\usepackage{grffile}
\graphicspath{{./fig/}}
\DeclareGraphicsExtensions{.pdf,.jpeg,.jpg,.png}
\usepackage{xcolor}
\usepackage{array}
\usepackage{multirow}
\usepackage{tabularx}
\usepackage{tabulary}
\usepackage{booktabs}
\usepackage{setspace}
\usepackage{lipsum}

\usepackage{soul}
\usepackage{etoolbox}
\makeatletter
\patchcmd{\SOUL@ulunderline}{\dimen@}{\SOUL@dimen}{}{}
\patchcmd{\SOUL@ulunderline}{\dimen@}{\SOUL@dimen}{}{}
\patchcmd{\SOUL@ulunderline}{\dimen@}{\SOUL@dimen}{}{}
\newdimen\SOUL@dimen
\makeatother

\usepackage{url}
\usepackage{hyperref}
\usepackage[plain]{fancyref}

\usepackage{tikz}
\usetikzlibrary{arrows.meta}
\usetikzlibrary{decorations.pathmorphing}
\usepackage[utf8]{inputenc}
\usepackage{pgfplots}
\DeclareUnicodeCharacter{2212}{−}
\usepgfplotslibrary{groupplots,dateplot}
\usetikzlibrary{patterns,shapes.arrows}
\pgfplotsset{compat=newest}

\newcommand{\isot}[2]{\(^{#2}\){#1}}

\DeclareMathOperator*{\diag}{diag}

\DeclareMathOperator*{\expect}{E}
\DeclareMathOperator*{\var}{var}

\pgfmathdeclarefunction{gauss}{2}{%
  \pgfmathparse{1/(#2*sqrt(2*pi))*exp(-((x-#1)^2)/(2*#2^2))}%
}

\pgfmathdeclarefunction{gausspoint}{3}{%
  \pgfmathparse{1/(#2*sqrt(2*pi))*exp(-((#3-#1)^2)/(2*#2^2))}%
}

\begin{document}
\IEEEpubid{\begin{minipage}{0.85\textwidth}\ \\[12pt]\\ \\ \\ \centering
  This work has been submitted to the IEEE for possible publication. Copyright may be transferred without notice, after which this version may no longer be accessible.
\end{minipage}}

\title{Mapping the Minimum Detectable Activities of Gamma-Ray Sources in a 3-D Scene}

\author{M.\,S.~Bandstra,
        D.~Hellfeld,
        J.~Lee,
        B.\,J.~Quiter,
        M.\,Salathe,  
        J.\,R.~Vavrek,
        and T.\,H.\,Y.~Joshi%
\thanks{M.\,S.~Bandstra, D.~Hellfeld, B.\,J.~Quiter, M.\,Salathe, J.\,R.~Vavrek, and T.\,H.\,Y.~Joshi are with the Nuclear Science Division at Lawrence Berkeley National Laboratory, Berkeley, CA 94720 USA e-mail: msbandstra@lbl.gov.}%
\thanks{J.~Lee is with the Nuclear Engineering Department at the University of California, Berkeley, Berkeley, CA 94720 USA.}%
\thanks{This material is based upon work supported by the Defense Threat Reduction Agency under HDTRA 10027--28018, 10027--30529.
This support does not constitute an express or implied endorsement on the part of the United States Government.
Distribution A: approved for public release, distribution is unlimited.}}

\maketitle


\begin{abstract}
The ability to formulate maps of minimal detectable activities (MDAs) that describe the sensitivity of an ad hoc measurement that used one or more freely moving radiation detector systems would be significantly beneficial for the conduct and understanding of many radiological search activities.
In a real-time scenario with a free-moving detector system, an MDA map can provide useful feedback to the operator about which areas have not been searched as thoroughly as others, thereby allowing the operator to prioritize future actions.
Similarly, such a calculation could be used to inform subsequent navigation decisions of autonomous platforms.
Here we describe a near real-time MDA mapping approach that can be applied when searching for point sources using detected events in a spectral region of interest while assuming a constant, unknown background rate.
We show the application of this MDA mapping method to a real scenario, a survey of the interior of a small building using a handheld detector system.
Repeated measurements with no sources and with \isot{Cs}{137} sources of different strengths yield results consistent with the estimated thresholds and MDA values; namely, that for background-only measurements no sources are seen above threshold anywhere in the scene, while when sources are present they are detected above the thresholds calculated for their locations.
\end{abstract}

\IEEEpeerreviewmaketitle%



\section{Introduction}
\IEEEPARstart{F}{inding} radioactive sources outside of regulatory control has been a major focus of research in recent decades.
Mobile, airborne, and human portable detector systems and detection algorithms have been developed to solve the problem of finding such sources amid the complex environments presented by urban scenes.

Many have studied the problem of localizing one or more point sources of radiation using a moving detector or an array of static detectors~\cite{howse_least_2001, nemzek_distributed_2004, stephens_detection_2004}, using a variety of methods, such as maximum likelihood~\cite{morelande_detection_2007, vilim_radtrac:_2009, deb_radioactive_2011, wan_detection_2012, deb_iterative_2013, baidoo-williams_theoretical_2015}, Bayesian methods~\cite{chandy_networked_2008, morelande_radiological_2009, ristic_information_2010, chin_efficient_2011, towler_radiation_2012, miller_adaptively_2015, rao_network_2015}, and geometric methods and clustering~\cite{chin_accurate_2008, rao_localization_2008, chin_identification_2010, wu_network_2019}.
The problem has been mostly explored by considering a two-dimensional (\mbox{2-D}) representation of the geometry and assuming bare point sources, but some have focused on \mbox{3-D} environments~\cite{sharma_three-dimensional_2016, bhattacharyya_estimating_2018, hellfeld_gamma-ray_2019, vavrek_reconstructing_2020} and considered attenuation in the scene~\cite{vilim_tracking_2009, hite_localization_2019, anderson_mobile_2020, bandstra_improved_2021}.

Recent advances in freely moving radiation detector systems that can sense their position and orientation in \mbox{3-D} space as well as map the \mbox{3-D} environment around them have enabled new search capabilities and prompted the development of new algorithms to fully exploit the contextual data available~\cite{barnowski_scene_2015, vetter_gamma-ray_2018, pavlovsky_3d_2019}.
One such approach is Point Source Likelihood (PSL), a maximum likelihood method in which photopeak events, detector positioning information, and the environmental map are used as inputs to an algorithm that attempts to locate a single point source in the \mbox{3-D} scene and obtain accurate estimates and confidence intervals for the source location and activity~\cite{hellfeld_gamma-ray_2019, vavrek_reconstructing_2020, hellfeld_free-moving_2021}, even in the presence of attenuating material~\cite{bandstra_improved_2021}.
A notable recent extension of PSL is solving for discrete Gaussian sources, allowing for truly continuous sources to be better fit~\cite{greiff_gamma-ray_2021}.

Despite the extensive study of point source localization, almost all studies have focused on detecting, localizing, and estimating the activity of point sources that are actually present near the detectors, although some have discussed the calculation of detection limits~\cite{wan_detection_2012}.
However, to our knowledge there are no approaches that have converted a detection limit into a map of MDA values in the scene in near-real time for arbitrary detector measurement configurations.

By contrast, making maps of MDA values (or, more typically, minimum detectable concentration (MDC) values) is standard in aerial surveys and \textit{in situ} measurements, where survey measurements are performed to search for contamination on or in the ground or water~\cite{nir-el_minimum_2001, bagatelas_determination_2010, tang_efficiency_2016}.
These surveys typically assume a planar geometry and fixed detector orientation during the survey portion of the flight, while a survey with a handheld instrument in a lab will involve a \mbox{3-D} geometry and more complex detector-source orientations and distances.

This work provides a method for extending existing methods in maximum likelihood point source search to mapping the MDA in a complex, \mbox{3-D} environment.
The focus is on small (order 100\,cm$^3$ active volume), handheld detector systems that can leverage their motion through the scene to find count-rate deviations that may provide contrary evidence to the null assumption that only a constant background is being measured.
The formulation is also relevant for larger detectors, but the assumption of constant background can be more difficult to satisfy.

The structure of this manuscript is as follows.
In~\Fref{sec:methods}, the existing PSL method is described and its extension to estimating MDA values is presented, including a method for estimating and accounting for the statistical correlations among PSL solutions, which strongly affect the estimated MDA values.
A comparison to the commonly used Currie formula for MDA is also made.
Then, in~\Fref{sec:toy-model}, a toy model is used to demonstrate the concepts presented in~\Fref{sec:methods}.
Finally, \Fref{sec:experimental-results} shows the application of the method to survey measurements with a handheld system, and~\Fref{sec:discuss} is a discussion of the results.


\section{Methods}\label{sec:methods}
Here we will introduce a maximum likelihood-based framework for analyzing the detection of point sources (Point Source Likelihood or PSL), and then we will use it to describe how to calculate the MDA\@.
Some of the notation and concepts used in this section will closely follow the description of PSL given in reference~\cite{bandstra_improved_2021}.


\subsection{Point Source Likelihood}\label{sec:psl}
The PSL algorithm is the reconstruction of a single point source using a freely moving detector in a \mbox{3-D} environment using maximum likelihood~\cite{hellfeld_gamma-ray_2019, vavrek_reconstructing_2020, bandstra_improved_2021, hellfeld_free-moving_2021, greiff_gamma-ray_2021}.
To perform PSL, we begin with a series of \(M\) measurements, indexed by~\(i\), that consist of event counts \(n_i\) in a chosen spectral region of interest (ROI).
The ROI is often a region around a photopeak; e.g., around the 662\,keV line of \isot{Cs}{137}.
In addition to the counts, the \mbox{3-D} locations~\(\mathbf{r}_i\) and orientations~\(\mathbf{q}_i\) of the detector at each measurement are determined, through, e.g., GPS locations and compass directions, or Simultaneous Localization and Mapping (SLAM)~\cite{durrant-whyte_simultaneous_2006, bailey_simultaneous_2006}.
Another ingredient needed for PSL is the detector response for the chosen spectral ROI and angle of incidence, expressed as the effective area~\(A\), which herein is the product of the geometric area and the full-energy detection efficiency.
The final requirement for PSL is a series of \textit{test points}, or hypothetical positions at which a point source may be present in the \mbox{3-D} environment.
We will assume there are \(N\) such points indexed by~\(j\).
Test points should be chosen carefully since they determine where the solution could lie, so they should be densely spaced and fill all relevant locations within the area of interest, but too many and the calculation could become too slow to perform in near-real time.
We typically use a 3-D grid of points throughout the search space with a pitch of 5--20\,cm depending on the size of the search space, and in general there will be many more points than measurements (\(N \gg M\)).
If available, some model of the \mbox{3-D} environment, e.g., a point cloud from a laser detection and ranging (LiDAR) unit, should be used to choose only points lying on or inside solid surfaces, since we assume the source or radioactive contamination has settled onto surfaces~\cite{bandstra_improved_2021, hellfeld_free-moving_2021}.

Each measurement is assumed to be the result of a Poisson process consisting of a constant background and a single possible point source at some location.
For a background count rate \(b\) and a source activity \(s\) at test point~\(j\), this assumption means
\begin{align}
    n_i &\sim \mathrm{Poisson}\left[\lambda_{ij}(b, s) \right], \\
    \lambda_{ij}(b, s) &\equiv (b + R_{ij} s) \Delta t_i,
\end{align}
where \(\lambda_{ij}\) is the mean of measurement~\(i\) due to a source at test point~\(j\), \(\Delta t_i\) is the integration time of measurement~\(i\), and \(R_{ij}\) is an element of the \(M \times N\) point-source response matrix \(\mathbf{R}\), which is
\begin{align}
    R_{ij} &= \frac{A(\mathbf{q}_i, \hat{\mathbf{r}}_{ij})}{4 \pi |\mathbf{r}_{ij}|^2} B C \tau_{ij},
\end{align}
Here \(\mathbf{r}_{ij} = \mathbf{r}_j - \mathbf{r}_i \) is the direction from the detector at measurement~\(i\) to the test point~\(j\) and \(\hat{\mathbf{r}}_{ij}\) is the corresponding unit vector.
The factor \(B\) is the photopeak branching ratio (e.g., \(B=0.85\) for the 662\,keV line of \isot{Cs}{137}), \(C\) is a factor to convert the desired source activity units into nuclear decays per second, and \(\tau_{ij}\) is the transmission factor for photons traveling along \(\mathbf{r}_{ij}\).
Here we assume no attenuation (\(\tau_{ij} = 1\)).

Maximum likelihood is used to simultaneously solve for the background count rate and source activity at each test point~\(j\).
Using the likelihood for Poisson-distributed data, the negative log-likelihood function minimized at test point~\(j\) is
\begin{align}
    -\log L_j(\mathbf{n} | b, s) &= \sum_{i=1}^{M} \left[\lambda_{ij}(b, s) - n_i \log \lambda_{ij}(b, s)\right], \label{eq:nll}
\end{align}
where terms that depend only on \(n_i\) have been dropped since the measurements are held constant during the iterative MLEM solution process.
The maximum likelihood expectation maximization (MLEM) multiplicative update rules~\cite{shepp_maximum_1982} can be used to solve for \(b\) and \(s\) starting from any nonzero initial guess, since the optimization problem is convex~\cite{bandstra_improved_2021}.
The resulting maximum likelihood estimates will be denoted \(\hat{b}_j\) and \(\hat{s}_j\).

PSL localizes and quantifies the activity of a point source by finding the test point index~\(j_{\mathrm{max}}\) whose fit results in the maximum likelihood over all the test points.
The likelihood-ratio test (LRT) and Fisher information are used to construct spatial and source activity confidence intervals, and the PSL algorithm can run in real time~\cite{bandstra_improved_2021}.


\subsection{Calculating the MDA for a single point}\label{sec:mda}
To estimate the minimum detectable activity (MDA) for a source at test point~\(j\), we first calculate the Fisher information matrix \(\mathbf{F}_j\) assuming the true values for \(b\) and \(s\) are known.
The Fisher information matrix is the expectation value of the Hessian of the negative log-likelihood function \(-\log L_j\), which is
\begin{align}
    \mathbf{F}_j(b, s) &= E \left\{ \left[ \begin{array}{cc}
        \sum_i \frac{n_i \Delta t_i^2}{\lambda_{ij}^2} & \sum_i \frac{n_i R_{ij} \Delta t_i^2}{\lambda_{ij}^2} \\
        \sum_i \frac{n_i R_{ij} \Delta t_i^2}{\lambda_{ij}^2} & \sum_i \frac{n_i R_{ij}^2 \Delta t_i^2}{\lambda_{ij}^2}
    \end{array} \right] \right\} \\
    &= \left[ \begin{array}{cc}
        \sum_i \frac{\Delta t_i}{b + R_{ij} s} & \sum_i \frac{R_{ij} \Delta t_i}{b + R_{ij} s} \\
        \sum_i \frac{R_{ij} \Delta t_i}{b + R_{ij} s} & \sum_i \frac{R_{ij}^2 \Delta t_i}{b + R_{ij} s}
    \end{array} \right] \\
    &\equiv \left[ \begin{array}{cc}
        (F_j)_{bb} & (F_j)_{bs} \\
        (F_j)_{bs} & (F_j)_{ss}
    \end{array} \right]
\end{align}
where we have defined the elements of \(\mathbf{F}_j\) for convenience.
Note that \(\mathbf{F}_j\) is symmetric by definition, and positive definite for this particular problem, since its two eigenvalues must be positive, as has been shown elsewhere~(e.g., \cite{bandstra_improved_2021}).

In maximum likelihood, the covariance matrix of the parameter estimates \(\hat{b}_j\) and \(\hat{s}_j\) approaches the inverse of the Fisher information matrix (i.e., the Cram\'{e}r-Rao lower bound (CRLB)) in the asymptotic limit, which means as a greater and greater total number of event counts are in the measurement~\cite{kendalls_advanced_statistics}.
In~\Fref{sec:validity} we will introduce criteria to ensure that we are close enough to the asymptotic case to use the CRLB as an approximation of the true covariance, although we will include an optional small multiplicative factor to allow for adjustments.
We define the following function to estimate the standard deviation of \(\hat{s}_j\) given the true background and source parameters \(b\) and \(s\):
\begin{align}
    \sigma_{\hat{s}_j}(b, s) &\equiv (1 + \eta) \sqrt{(F_j)^{-1}_{ss}} \label{eq:sigma_s} \\
    &= (1 + \eta) \sqrt{\frac{(F_j)_{bb}}{(F_j)_{bb} (F_j)_{ss} - (F_j)_{bs}^2}}, \label{eq:sigma-s-hat}
\end{align}
where \(\eta \ll 1\) is a small nonnegative factor to allow for adjustments for non-asymptotic behavior.
(To first order, \(\eta \sim 1 / \sum_i n_i\)~\cite{kendalls_advanced_statistics}.)
The exact value of \(\eta\) will depend on the mean count rate and duration of the measurement, and for a particular system under typical operational scenarios the value could be adjusted until a desired false positive rate is achieved.

If \(s = 0\) and there were no nonnegativity constraints on \(\hat{s}_j\), one expects under maximum likelihood that in the asymptotic limit \(\hat{s}_j\) would be distributed as a Gaussian with mean of zero and standard deviation \(\sigma_{\hat{s}_j}(b, 0) \propto \sqrt{b}\).
However, imposing a nonnegativity constraint has the effect of clipping the values of \(\hat{s}_j\) at zero when they would have been negative, and so then asymptotically \(\hat{s}_j\) is distributed according to a rectified Gaussian of the same parameters.
Since we will only be concerned with its cumulative distribution function (CDF) going forward, this clipping of the distribution will not affect the analysis but it will be apparent in histograms of \(\hat{s}_j\).

Following the formulation of Currie~\cite{currie_limits_1968}, to estimate the MDA we choose a false positive probability \(\alpha\) under background conditions (\(s = 0\)) and false negative probability \(\beta\) under source conditions (\(s > 0\)).
The common choice of \(\alpha = \beta = 0.05\) was used here.

\begin{figure}
\begin{center}
\begin{tikzpicture}
\pgfmathsetmacro{\smax}{7}
\pgfmathsetmacro{\sigmazero}{1}
\pgfmathsetmacro{\sigmamda}{1.4}
\pgfmathsetmacro{\scrit}{1.645 * \sigmazero}
\pgfmathsetmacro{\smda}{\scrit + 1.645 * \sigmamda}

\definecolor{matplotlibblue}{RGB}{58,118,175}
\definecolor{matplotlibred}{RGB}{197,57,50}

\begin{axis}[
    no markers,
    domain=0:\smax,
    ymin=0,
    ymax={1.08 * gausspoint(0, \sigmazero, 0)},
    xmin=0,
    xmax={\smax+0.1},
    samples=100,
    axis lines=center,
    axis line style = thick,
    every axis x label/.style={at=(current axis.right of origin),anchor=west},
    height=6cm, width=9.5cm,
    xtick={0, \scrit, \smda},
    xticklabels={0, $s_j^{\mathrm{crit}}$, $s_j^{\mathrm{mda}}$},
    xtick=\empty,
    ytick=\empty,
    enlargelimits=false, clip=false, axis on top,
    grid = major
    ]
    \addplot [fill=matplotlibblue!20, draw=none, domain=0:\smax] {gauss(0,\sigmazero)} \closedcycle;
    \addplot [fill=matplotlibred!20, draw=none, domain=0:\smax] {gauss(\smda,\sigmamda)} \closedcycle;
    \addplot [fill=matplotlibblue!60, draw=none, domain=\scrit:\smax] {gauss(0,\sigmazero)} \closedcycle;
    \addplot [fill=matplotlibred!60, draw=none, domain=0:\scrit] {gauss(\smda,\sigmamda)} \closedcycle;
    \addplot [very thick,matplotlibblue!80!black, opacity=0.7, domain=0:\smax] {gauss(0,\sigmazero)};
    \addplot [very thick,matplotlibred!80!black, opacity=0.7] {gauss(\smda,\sigmamda)};

    \node at (axis cs: 0, {0.5 * gausspoint(0, \sigmazero, 0)}) [xshift=-0.4cm, rotate=90, text centered, anchor=center] {$P(\hat{s}_j)$};
    \node at (axis cs: 0, 0) [yshift=-0.3cm, text centered, anchor=center] {$0$};
    \node at (axis cs: {\smax}, 0) [xshift=0.1cm, yshift=-0.3cm, text centered, anchor=center] {$\hat{s}_j$};
    \draw [color=black, dotted, thick] (axis cs: \scrit, {gausspoint(0, \sigmazero, \scrit)}) -- (axis cs: \scrit, 0) node [yshift=-0.3cm, text centered, anchor=center] {$s_j^{\mathrm{crit}}$};

    \draw [{Latex}-{Latex}, color=matplotlibblue!50!black] (axis cs: 0, {gausspoint(0, \sigmazero, \sigmazero)}) -- (axis cs: \sigmazero, {gausspoint(0, \sigmazero, \sigmazero)}) node [midway, xshift=0.1cm, yshift=-0.3cm, text centered, anchor=center] {\footnotesize $\sigma_{\hat{s}_j}(b, 0)$};
    \draw [-, color=matplotlibblue!50!black] (axis cs: {\scrit+0.6}, 0.07) node [xshift=0.2cm, yshift=0.2cm, text centered, anchor=center] {$\alpha$} -- (axis cs: {\scrit+0.15}, 0.02);

    \draw [{Latex}-{Latex}, color=matplotlibred!50!black] (axis cs: {\smda-\sigmamda}, {gausspoint(0, \sigmamda, \sigmamda)}) -- (axis cs: \smda, {gausspoint(0, \sigmamda, \sigmamda)}) node [midway, xshift=-0.1cm, yshift=-0.3cm, text centered, anchor=center] {\footnotesize $\sigma_{\hat{s}_j}(b, s_j^{\mathrm{mda}})$};
    \draw [-, color=matplotlibred!50!black] (axis cs: {\scrit-0.6}, 0.07) node [xshift=-0.2cm, yshift=0.2cm, text centered, anchor=center] {$\beta$} -- (axis cs: {\scrit-0.15}, 0.02);
    \draw [color=black, dotted, thick] (axis cs: \smda, {gausspoint(\smda, \sigmamda, \smda)}) -- (axis cs: \smda, 0) node [yshift=-0.3cm, text centered, anchor=center] {$s_j^{\mathrm{mda}}$};
\end{axis}
\end{tikzpicture}
\end{center}
\caption{Demonstration of how MDA is calculated for a single test point, inspired by Figure~2 from~\cite{currie_limits_1968}.
Not pictured are the delta functions at \(\hat{s}_j=0\) that result from the non-negativity constraint.\label{fig:mda_diagram}}
\end{figure}
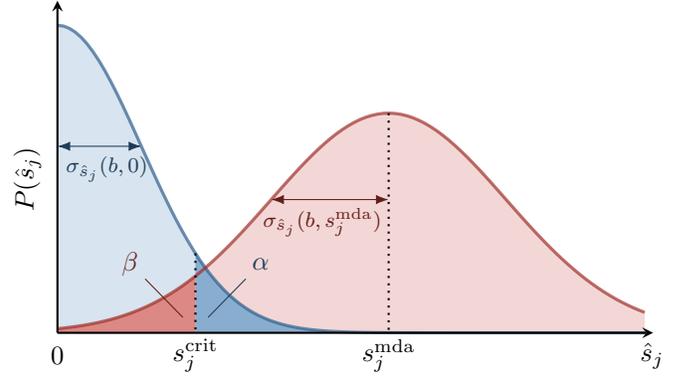

The critical value, i.e., the source detection threshold, is defined as
\begin{align}
    s_j^{\mathrm{crit}} &= k_{\alpha}\, \sigma_{\hat{s}_j}(b, 0) \propto \sqrt{b},
\end{align}
where \(k_{\alpha} \equiv \Phi^{-1}(1 - \alpha)\) and \(\Phi\) is the cumulative distribution function (CDF) of the standard normal distribution.
The MDA is found by increasing \(s\) until the following condition is satisfied at the value \(s_j^{\mathrm{mda}}\):
\begin{align}
    s_j^{\mathrm{mda}} = s_j^{\mathrm{crit}} + k_{\beta}\, \sigma_{\hat{s}_j}(b, s_j^{\mathrm{mda}}), \label{eq:s_mda}
\end{align}
which is the activity at which a set of measurements of constant background rate \(b\) would be able to detect a point source with a confidence level of \(1 - \beta\) and reject background at a confidence level of \(1 - \alpha\).
For \(\alpha = \beta = 0.05\), \(k_{\alpha} = k_{\beta} \approx 1.645\).
Since \fref{eq:s_mda} cannot in general be solved in closed form, an iterative method such as a bisection search must be used.
\Fref{fig:mda_diagram} shows a diagram of the relationship between the critical value and MDA\@.

\subsection{Estimating a correction to account for all test points}\label{sec:multiple-point-correction}
In~\Fref{sec:mda} it was assumed that only a single statistical test was relevant (i.e., the presence or absence of a point source at a single test point); however, since PSL may potentially consider thousands of test points simultaneously, if the critical value from the previous section were used as-is then the false positive rate for a detection of a source at \textit{any} test point would be larger than~\(\alpha\), even approaching 100\%.
To tackle this problem, here we will present a simplified version of PSL, analyze it for insights into the statistical correlations of its solutions, and use this analysis to derive an approximate correction to \(k_{\alpha}\).

To analyze the problem, we will temporarily adopt the Gaussian approximation to simplify PSL, so instead of assuming the counts have a Poisson distribution (\(n_i \sim \mathrm{Poisson}(b \Delta t_i)\)), for the moment we will instead assume the counts are drawn from a Gaussian distribution with the mean and variance of the Poisson distribution, i.e.,
\begin{align}
    n_i &\sim {\cal N}(b \Delta t_i, b \Delta t_i).
\end{align}
Minimizing the log likelihood function then takes the form of a least squares problem:
\begin{align}
    - \log L_j(\mathbf{n} | b_j, s_j) &= \sum_{i=0}^{M-1} \frac{[ n_i - (b_j + s_j R_{ij}) \Delta t_i ]^2}{2 b \Delta t_i}.
\end{align}
This function can be straightforwardly minimized by setting the gradients with respect to both \(b_j\) and \(s_j\) to zero.
The resulting minimum least-squares estimator for \(s_j\) can be written as a vector dot product
\begin{align}
    \hat{s}_j &= \frac{\left( \sum_i \Delta t_i \right) \mathbf{R}_{j} - \left( \sum_i R_{ij} \Delta t_i \right) \mathbf{1}_M}{\left( \sum_i \Delta t_i \right) \left( \sum_i R^2_{ij} \Delta t_i \right) - \left( \sum_i R_{ij} \Delta t_i \right)^2} \cdot \mathbf{n} \\
    &\equiv \mathbf{W}_{j} \cdot \mathbf{n}
\end{align}
where \(\mathbf{R}_{j}\) is the \(j\)th column of \(\mathbf{R}\) and \(\mathbf{1}_M\) is a column vector of \(M\) ones.
The vector \(\mathbf{W}_j\) is the \(j\)th column vector of the \(M \times N\) source activity estimation matrix \(\mathbf{W}\), making the vector of least squares source estimates for \textit{all} points
\begin{align}
    \hat{\mathbf{s}} &= \mathbf{W}^{\top} \mathbf{n}. \label{eq:shat-WT-n}
\end{align}

Each of these estimated source activities is unbiased, i.e.,
\begin{align}
    \expect[\hat{s}_j] &= \mathbf{W}_j \cdot \expect[\mathbf{n}] = \mathbf{W}_j \cdot b \mathbf{\Delta t} \\
    &\propto \left( \sum_i \Delta t_i \right) \mathbf{R}_{j} \cdot \mathbf{\Delta t} - \left( \sum_i R_{ij} \Delta t_i \right) \mathbf{1}_M \cdot \mathbf{\Delta t} \\
    &= 0, \label{eq:source-mean-zero}
\end{align}
so
\begin{align}
    \expect[\hat{\mathbf{s}}] &= \mathbf{0}_N,
\end{align}
where \(\mathbf{0}_N\) is a column vector of \(N\) zeros.

In this approximation, \(\hat{\mathbf{s}}\) is distributed as a multivariate normal with a mean of \(\mathbf{0}_N\).
Its covariance matrix is
\begin{align}
    \var[\hat{\mathbf{s}}] &= \mathbf{W}^{\top} \var[\mathbf{n}]\, \mathbf{W} = \mathbf{W}^{\top} \diag(b \mathbf{\Delta t})\, \mathbf{W}
\end{align}
and the variances along the diagonal can be shown to be the same as those obtained from the Fisher information matrix in \fref{eq:sigma_s}, i.e.,
\begin{align}
    \left(\var[\hat{\mathbf{s}}]\right)_{jj} &= \var[\hat{s}_j] = \sigma^2_{\hat{s}_j}(b, 0),
\end{align}
which is not surprising because the asymptotic limit is assumed for both this analysis and the analysis leading to \fref{eq:sigma_s}.

The following singular value decomposition (SVD) will be helpful in diagonalizing \(\var[\hat{\mathbf{s}}]\) so that we can find a transformation to uncorrelated variables:
\begin{align}
    \diag\left(\sqrt{b \mathbf{\Delta t}} \right)\, \mathbf{W} &= \mathbf{U} \boldsymbol\Sigma \mathbf{V}^{\top},
\end{align}
where \(\mathbf{U}\) is an \(M \times M\) unitary matrix, \(\boldsymbol\Sigma\) is an \(M \times N\) diagonal matrix, and \(\mathbf{V}\) is an \(N \times N\) unitary matrix.
Because we can assume in practice that we can always have more points than measurements (\(N > M\)), we can immediately remove all but the first \(M\) columns of both \(\boldsymbol\Sigma\) and \(\mathbf{V}\) and preserve the exact equality.
Then \(\boldsymbol\Sigma\) is an \(M \times M\) square diagonal matrix, and \(\mathbf{V}\) is an \(N \times M\) matrix that is no longer unitary but has orthonormal columns: \(\mathbf{V}^{\top} \mathbf{V} = \mathbf{I}_M\), but \(\mathbf{V} \mathbf{V}^{\top} \neq \mathbf{I}_N\).
It should be noted that SVD can be efficiently calculated when limiting the number of columns of \(\mathbf{V}\) to \(M\) instead of (the potentially orders of magnitude larger) \(N\) by using algorithms such as \texttt{TruncatedSVD} available in \texttt{scikit-learn}~\cite{scikit-learn}.

The SVD decomposition diagonalizes \(\var[\hat{\mathbf{s}}]\) in the following way:
\begin{align}
    \var[\hat{\mathbf{s}}] &= \mathbf{W}^{\top} \diag(b \mathbf{\Delta t})\, \mathbf{W} \\
    &= \left[ \diag\left(\sqrt{b \mathbf{\Delta t}} \right)\, \mathbf{W} \right]^{\top} \left[ \diag\left(\sqrt{b \mathbf{\Delta t}} \right)\, \mathbf{W} \right] \\
    &= \left( \mathbf{U} \boldsymbol\Sigma \mathbf{V}^{\top} \right)^{\top} \left( \mathbf{U} \boldsymbol\Sigma \mathbf{V}^{\top} \right) \\
    &= \mathbf{V} \boldsymbol\Sigma \mathbf{U}^{\top} \mathbf{U} \boldsymbol\Sigma \mathbf{V}^{\top} \\
    &= \mathbf{V} \boldsymbol\Sigma \mathbf{I}_M \boldsymbol\Sigma \mathbf{V}^{\top} \\
    &= \mathbf{V} \boldsymbol\Sigma^2 \mathbf{V}^{\top}
\end{align}

We can then transform \(\hat{\mathbf{s}}\) into orthonormal eigenmodes \(\mathbf{z}\):
\begin{align}
    \mathbf{z} &\equiv \boldsymbol\Sigma^{-1} \mathbf{V}^{\top} \hat{\mathbf{s}}. \label{eq:transform_s_to_z}
\end{align}
We can show that each dimension of \(\mathbf{z}\) is statistically independent and has a unit normal distribution.
To be unit normal, the means must all be zero:
\begin{align}
    \expect[\mathbf{z}] &= \boldsymbol\Sigma^{-1} \mathbf{V}^{\top} \expect[\hat{\mathbf{s}}] = \mathbf{0}_N
\end{align}
and the covariance matrix is the identity matrix:
\begin{align}
    \var[\mathbf{z}] &= \boldsymbol\Sigma^{-1} \mathbf{V}^{\top} \var[\hat{\mathbf{s}}] \mathbf{V} \boldsymbol\Sigma^{-1} \\
    &= \boldsymbol\Sigma^{-1} \mathbf{V}^{\top} \mathbf{V} \boldsymbol\Sigma^2 \mathbf{V}^{\top} \mathbf{V} \boldsymbol\Sigma^{-1} \\
    &= \boldsymbol\Sigma^{-1} \mathbf{I}_M \boldsymbol\Sigma^2 \mathbf{I}_M \boldsymbol\Sigma^{-1} \\
    &= \mathbf{I}_M.
\end{align}

In other words, under the Gaussian approximation considered in this section, we have transformed from \(N\) correlated multivariate-normal random variables (\(\hat{\mathbf{s}}\)) into \(M\) statistically independent unit-normal random variables (\(\mathbf{z}\)).
Statistical fluctuations in the background will ``excite'' these modes in proportion to the size of their eigenvalues, in an analogous way to the excitation of vibrational modes in acoustic and mechanical systems.
By estimating how large the several strongest modes are on average, we can estimate how large of an ``envelope'' is needed to describe these collective fluctuations at each test point, and therefore increase the detection thresholds beyond what a single test point would require on its own.

For purposes here, we are interested in how these modes approximate the statistical correlations of the PSL solutions across all test points, which will inform how we can adjust the scaling factor \(k_{\alpha}\) for the critical value upwards from the single-point estimate presented earlier, which was \(k_{\alpha} = \Phi^{-1}(1 - \alpha)\).
For this final step, we first consider only the first \(K\) dimensions of \(\mathbf{z}\) such that the fraction of the explained variance of those \(K\) eigenmodes (\(\sum_{j=0}^{K-1} \Sigma_{jj}^2 / \sum_{j=0}^{M-1} \Sigma_{jj}^2\)) is at least 99\%, which was selected arbitrarily.
(This selection was made to speed up the next step by neglecting any eigenmodes that do not contribute a significant amount of variability to \(\hat{\mathbf{s}}\).)
Then, a large number \(J\) of random samples of \(\mathbf{z}\) are made, amounting to \(J K\) samples from a unit normal distribution.
The last \(M-K\) elements of each \(\mathbf{z}_{\mathrm{sample}}\) can either be set to zero or the relevant matrices be further truncated.
These samples are projected into \(\hat{\mathbf{s}}\) values using the pseudoinverse of~\fref{eq:transform_s_to_z}:
\begin{align}
    \hat{\mathbf{s}}_{\mathrm{sample}} &= \mathbf{V} \boldsymbol\Sigma\, \mathbf{z}_{\mathrm{sample}}.
\end{align}
Finally, \(\sigma_{\hat{s}_j}(b, 0)\) can be calculated for each test point according to~\fref{eq:sigma_s}, denoted \(\boldsymbol\sigma(b, 0)\) in vectorized form.
The scaling factor for each sample can be calculated as \( k = \max \| \hat{\mathbf{s}}_{\mathrm{sample}} / \boldsymbol\sigma(b, 0) \| \), and \(k_{\alpha}\) was determined by finding the value of \(k\) such that a fraction of \(1 - 2\alpha\) sample \(k\) values were below \(k_{\alpha}\) and a fraction of \(2 \alpha\) were above it.
The absolute value could be used since the reconstructed values of \(\hat{\mathbf{s}}\) are symmetric around zero, and thus the samples can be ``reused'' for greater statistical power, while the factor of 2 is needed to compensate for this effect.

For more clarity and an example of calculating this correction, please see~\Fref{sec:toy-model} where a specific example is worked through, including plots of some of the eigenmodes to see how strongly the PSL solutions are correlated among the test points.

As a final note, \fref{eq:shat-WT-n} could also be sampled directly without performing SVD, and in some cases this approach might be less complicated without greatly increasing calculation time.
In this approach, \(\mathbf{z}_{\mathrm{sample}}\) has length \(M\) and each element is sampled from a unit normal, so then each estimated vector of source activities is
\begin{align}
    \hat{\mathbf{s}}_{\mathrm{sample}} &= \mathbf{W}^{\top} \left[ b \mathbf{\Delta t} + \mathrm{diag}\left(\sqrt{b \mathbf{\Delta t}}\right) \mathbf{z}_{\mathrm{sample}} \right] \\
    &= \mathbf{W}^{\top} \mathrm{diag}\left(\sqrt{b \mathbf{\Delta t}}\right) \mathbf{z}_{\mathrm{sample}}.
\end{align}
Therefore \(J M\) unit normal samples would be required, and, unless \(K \ll M\), this more direct sampling approach might be preferred.


\subsection{Requirements for a valid MDA approximation}\label{sec:validity}
To this point it has been assumed that the true background and source values were known, however in actual measurements they will not be known.
The only information available about the background and source is what can be gleaned from the measurements themselves, and that information will be limited by the total number of events detected.
The limited information provided by the measurements leads to two requirements for the MDA approximation to be valid.

One requirement comes from estimating how precisely the background is known assuming no source is present.
Assuming the simplest case that the true source activity is zero, minimizing~\fref{eq:nll} leads to the mean background value
\begin{align}
    \bar{b} &\equiv \frac{\sum_i n_i}{\sum_i \Delta t_i},
\end{align}
which, as expected, has no dependence on test point index~\(j\).
The standard deviation of \(\bar{b}\) can be estimated as \(1 / \sqrt{(F_j)_{bb}}\), which reduces to
\begin{align}
    \sigma_{\bar{b}} &= \sqrt{\frac{\bar{b}}{\sum_i \Delta t_i}} \equiv \frac{\sqrt{\sum_i n_i}}{\sum_i \Delta t_i},
\end{align}
and which is what would be expected from the Poisson nature of \(\sum_i n_i\).

In order to use \(\bar{b}\) in the place of the true background \(b\), it is useful to require that the relative uncertainty of \(\bar{b}\) is a small fraction of \(\bar{b}\), e.g.,
\begin{align}
    \frac{\bar{b}}{\sigma_{\bar{b}}} &\ge k_b \label{eq:require_b_bar}
\end{align}
where \(k_b\) is the number of ``sigma'' that \(\bar{b}\) is away from zero.
Choosing \(k_b = 10\) as a threshold for using the approximation leads to the requirement that
\begin{align}
    \sum_i n_i &\ge k_b^2 = 100, \label{eq:require_total_counts_gt_kb2}
\end{align}
or, in other words, that the total number of events is large enough to use the Gaussian approximation to the Poisson distribution.
This condition not only ensures that we can assume \(\bar{b} \approx b\), but also that the statistically asymptotic relationship implicit in~\fref{eq:sigma_s} is a good, though not perfect, approximation.

However, if a point source is actually present, using \(\bar{b}\) as the background estimate in the MDA calculation can lead to MDA values that are biased higher than they need to be, since the additional counts from the source would increase \(\bar{b}\) above the true background rate, and, as shown earlier, \(s_j^{\mathrm{crit}} \propto \sqrt{\bar{b}}\).
For a weak source the effect may not matter, but for a strong source it will lead to MDA estimates that are larger than necessary.
It is possible within this framework to account for the possible presence of a point source before calculating the MDA, by, e.g., using the fitted background at the best fit point~\(\hat{b}_{j_{\mathrm{max}}}\) in place of \(\bar{b}\).
However, the added complexity of this approach, coupled with the possibility of model mismatch, probably severely limits its utility.


\subsection{Relationship to the Currie formula}
The formulation of the previous sections provides a way to estimate the MDA for the general case when the background and source cannot be cleanly separated in time.
The well known Currie formula for radioactive signal detection for paired observations (i.e., background only versus background with source)~\cite{currie_limits_1968} can be recovered for the special case when \(R_{i} = 0\) for half of the measurement time and \(R_{i} = R\) is constant for the other half of the measurement time.
(There is effectively only a single test point in this situation, so we have dropped the index~\(j\) for the time being, and we do not need to apply the correction to \(k_{\alpha}\) described in~\Fref{sec:multiple-point-correction}.)

Letting \(T\) be half of the measurement time, then the elements of the Fisher information matrix are
\begin{align}
    F_{bb} &= \frac{T}{b} + \frac{T}{b + R s} \\
    F_{bs} &= \frac{R T}{b + R s} \\
    F_{ss} &= \frac{R^2 T}{b + R s}.
\end{align}
Then we get
\begin{align}
    \sigma_{\hat{s}}(b, s) &= (1 + \eta) \frac{\sqrt{2 b T + R s T}}{R T}
\end{align}
Assuming \(\alpha = \beta\) and \(\eta = 0\), this formula leads to the critical value
\begin{align}
    s^{\mathrm{crit}} &= k_{\alpha} \frac{\sqrt{2 b T}}{R T}
\end{align}
and so
\begin{align}
    s^{\mathrm{mda}} &= s^{\mathrm{crit}} + k_{\alpha} \frac{\sqrt{2 b T + R s^{\mathrm{mda}} T}}{R T}.
\end{align}
Finally, solving for \(s^{\mathrm{mda}}\) yields the familiar formula for paired background and source observations~\cite{currie_limits_1968}:
\begin{align}
    s^{\mathrm{mda}} &= \frac{k_{\alpha}^2 + 2 \sqrt{2}\, k_{\alpha} \sqrt{b T}}{R T} \\
    &\approx \frac{2.71 + 4.65 \sqrt{b T}}{R T}
\end{align}
once again using \(\alpha = 0.05\) and the single-point value for \(k_{\alpha}\) of \(1.645\) and assuming \(\eta \rightarrow 0\).
The numerator contains \(b T\), which is the mean number of counts due to background during each half of the measurement.
The denominator \(R T\) is the sensitivity to the test point, i.e., the mean number of counts due to a source being present per unit source activity, and so \(R s^{\mathrm{mda}} T\) is the expected number of counts in addition to background due to the source at the MDA\@.

Finally, the \(\bar{b}\) validity condition~\fref{eq:require_b_bar} with \(k_b = 10\) leads to the requirement that \(\sum_i n_i \ge 100\).
Since the mean of \(\sum_i n_i\) is \(2 b T\), this requirement translates to \(b T \ge 50\), which is slightly more stringent than the common requirement that \(b T > 30\) for using the Gaussian approximation to the Poisson distribution (e.g.,~\cite{knoll_radiation_2010}).


\subsection{Application of the MDA approach to other situations}
PSL and the MDA calculation have been presented assuming a single detector system with a single detector channel, but the approach can be straightforwardly extended to multiple detector modules and/or multiple detector systems.
To extend PSL, one must decide whether to require that all detector channels have the same background rate, have known channel-specific fixed proportions, or have rates that must be fit independently of one another.
To extend the MDA calculation, the situation is simpler --- one can just use the most appropriate background estimate \(\bar{b}\) for whichever detector channel a given measurement~\(i\) pertains to.
Thus the MDA method shown here can be applied without much modification to one or more detector systems, each of which may or may not be moving, and each of which may contain one or more detector modules.

Another case is when the background level is to be regarded as known.
In this case, the standard deviation of the estimator \(\hat{s}_j\) (\fref{eq:sigma_s}) reduces to
\begin{align}
    \sigma_{\hat{s}_j}(b, s) &= (1 + \eta) \frac{1}{\sqrt{(F_j)_{ss}}} \\
    &= \frac{1 + \eta}{\sqrt{\sum_i \frac{R_{ij}^2 \Delta t_i}{b + R_{ij} s}}}
\end{align}
and it would be prudent to also perform a qualitative or quantitative test that the measured background is indeed in agreement with the presumed background rate.


\section{Results for a toy problem}\label{sec:toy-model}
\begin{figure}
\begin{center}
    \includegraphics[width=0.99\columnwidth]{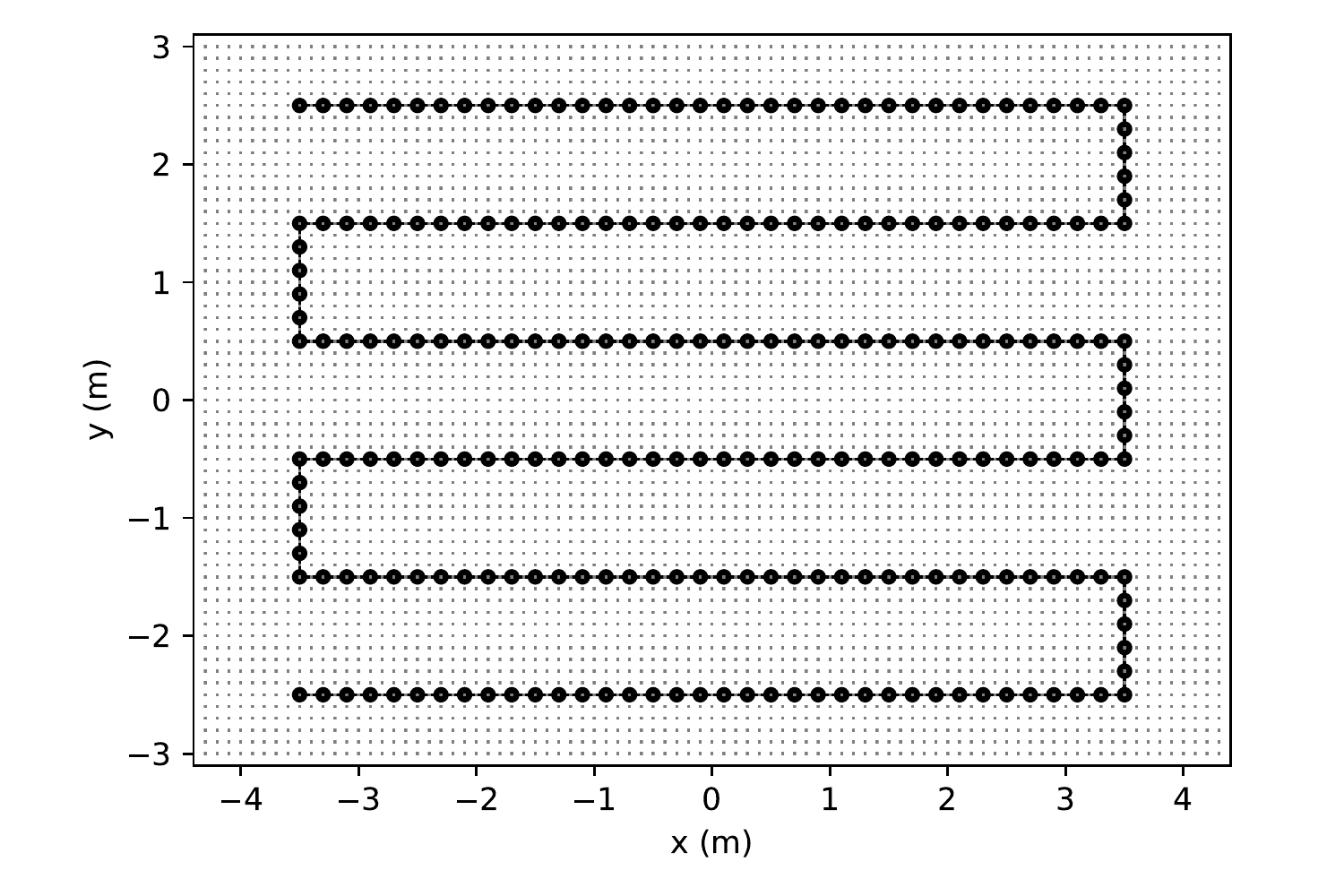}
\end{center}
\caption{Top-down view of the measurement path (bold black points) and test points (small gray points) used in the toy problem.
The measurement points are in a plane one meter above the test points.\label{fig:toy_path}}
\end{figure}

To demonstrate the MDA approximation method outlined in~\Fref{sec:methods}, a simple toy problem was devised.
In the scenario a small, handheld detector traversed a search area in a raster pattern.
The detector was assumed to have an isotropic effective area of 5\,cm\(^2\) for photons from a hypothetical gamma-transition with a branching ratio of~100\%.
The detector was moved in a raster pattern at a speed of 0.4\,m/s, which was approximated as a series of discrete measurement points spaced along a path by 0.2\,m, each point with a dwell time of 0.5\,s, and the simulated mean background rate was 2.0\,cps.
In total, 236 discrete measurement positions were simulated, comprising a  measurement duration of 118\,s.
The test points were chosen to lie in a plane 1\,m below the plane of the raster to simulate a ``floor'' over which the search was taking place, and the points were chosen to lie on a regularly spaced grid with a pitch of 10\,cm, for a total of 5,307 points.
The exact arrangements of the measurement locations and the test points are shown in~\Fref{fig:toy_path}.

\begin{figure*}
\begin{center}
    \includegraphics[width=0.42\textwidth]{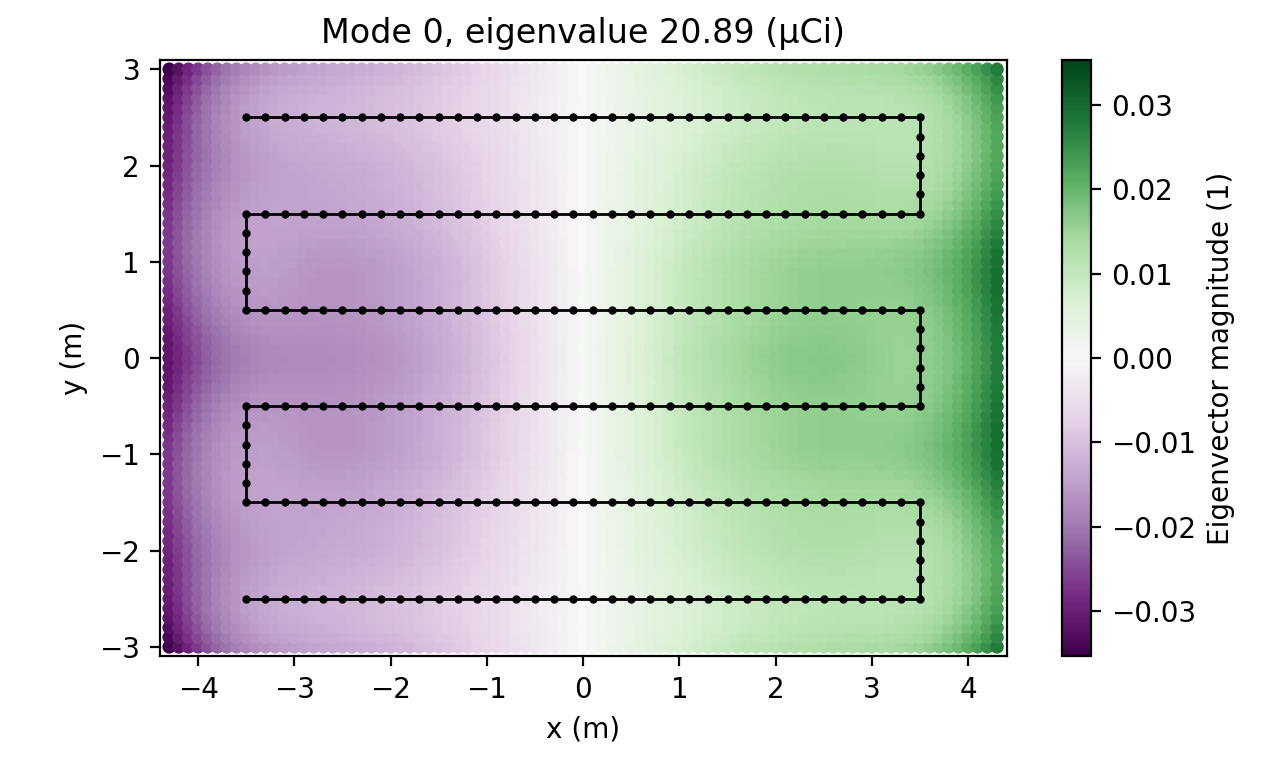}
    \includegraphics[width=0.42\textwidth]{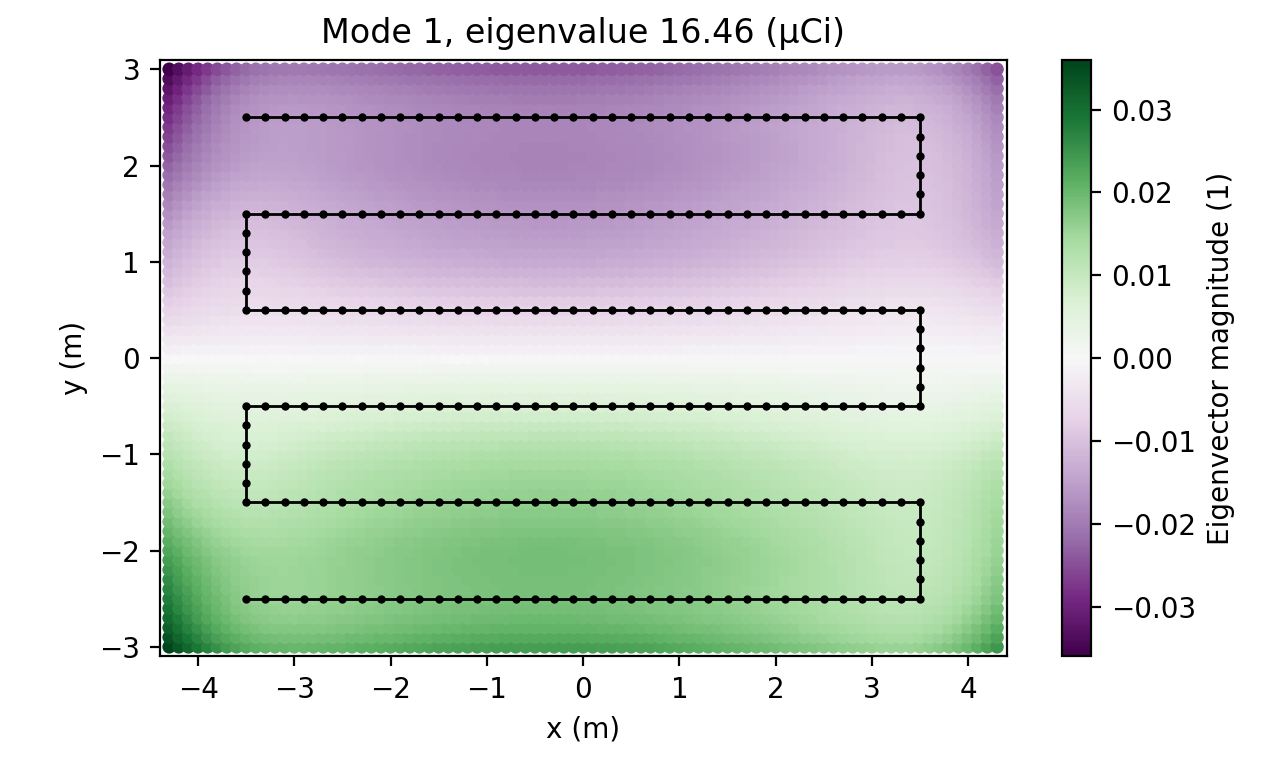} \\
    \includegraphics[width=0.42\textwidth]{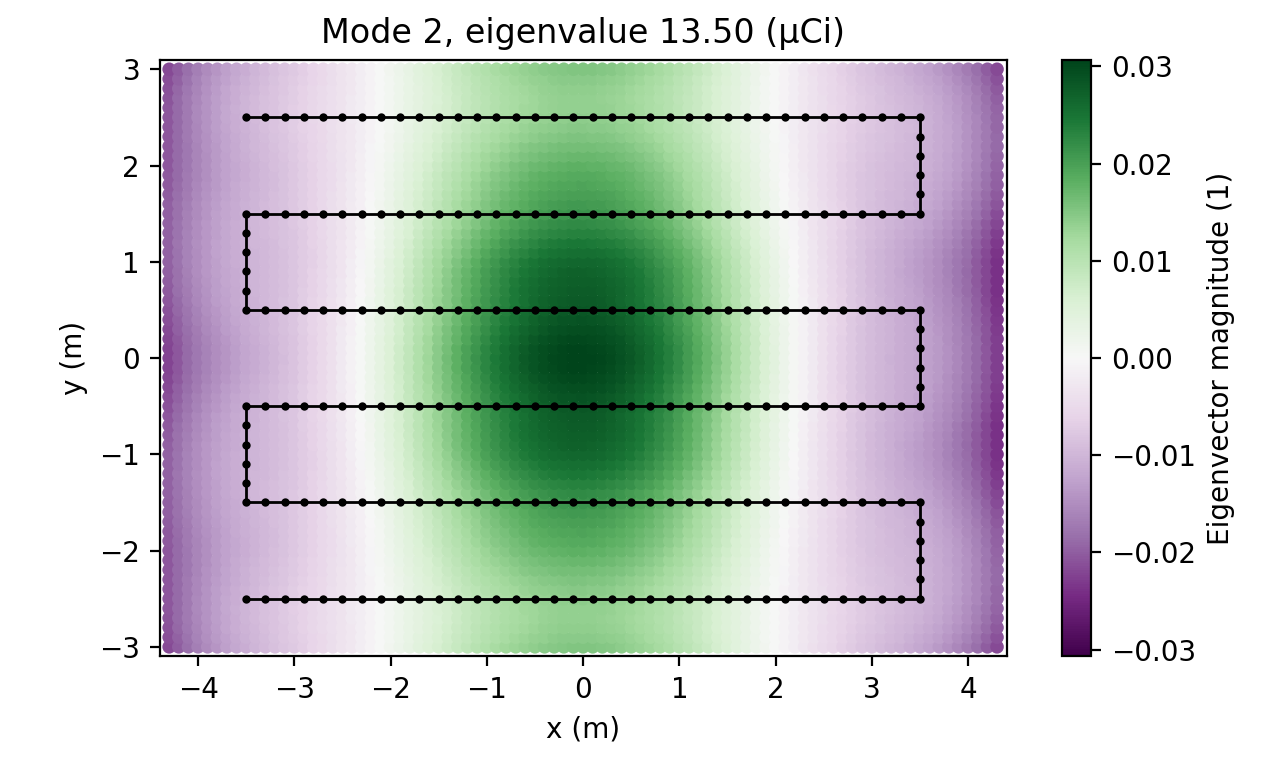}
    \includegraphics[width=0.42\textwidth]{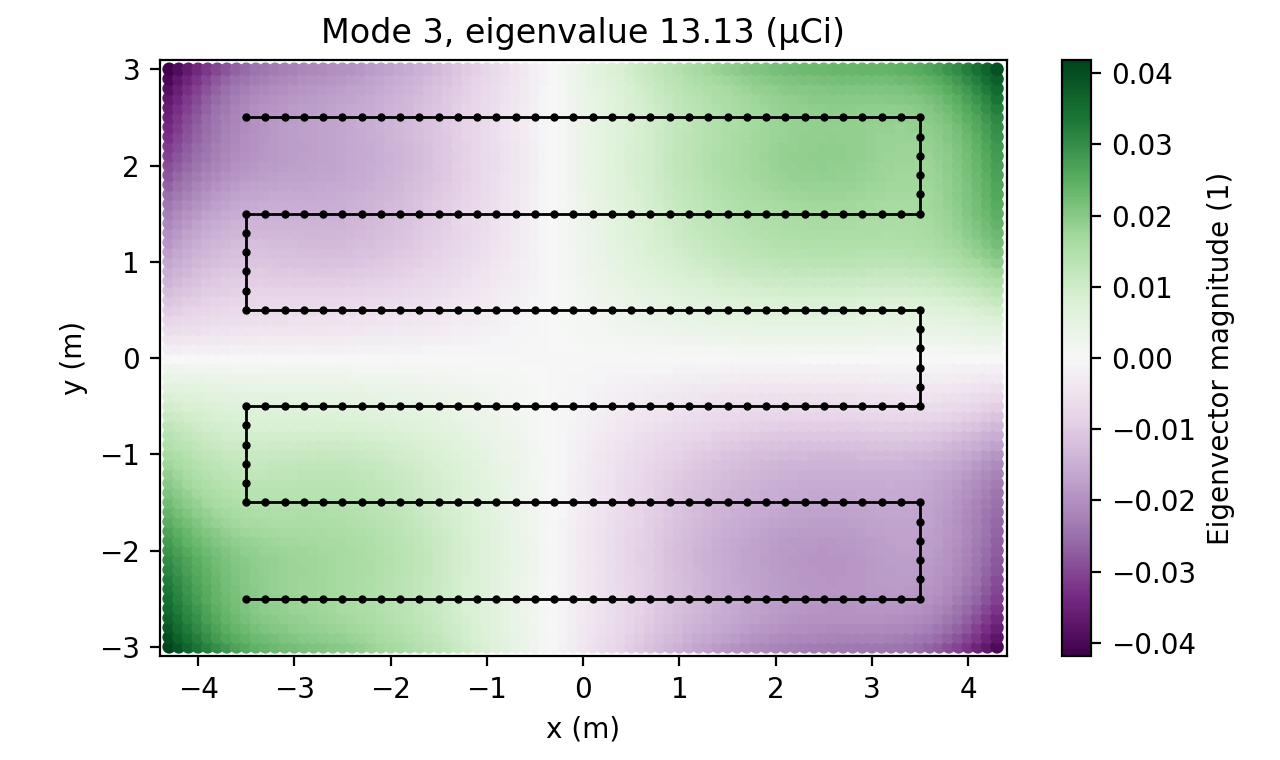}
\end{center}
\caption{The first four eigenmodes of the source reconstruction matrix for the toy model assuming a background count rate of \(b\) of 2\,cps.
These modes are excited by random samples of the measurements and represent correlations between the test point solutions.
These correlations must be accounted for when estimating the critical values (detection thresholds) at each point.\label{fig:toy_modes}}
\end{figure*}

The correction described in~\Fref{sec:multiple-point-correction} was calculated for the toy scenario.
The first four eigenmodes found by the method are shown in~\Fref{fig:toy_modes}.
Using a cutoff of 99\% on explained variance, only the first \(K\)~=~31 eigenmodes were needed, instead of the maximum of \(M\)~=~236 modes.
After generating \(J\)~=~10,000 random samples (i.e., a total of \(J K\)~=~310,000 draws from a unit normal distribution), the correction factor was found.
The final result was an increase in \(k_{\alpha}\) from 1.645 (assuming \(\alpha\)~=~0.05 and a single point) to 3.023 to account for the statistical correlations between the test points.

\begin{figure}
\begin{center}
    \includegraphics[width=0.99\columnwidth]{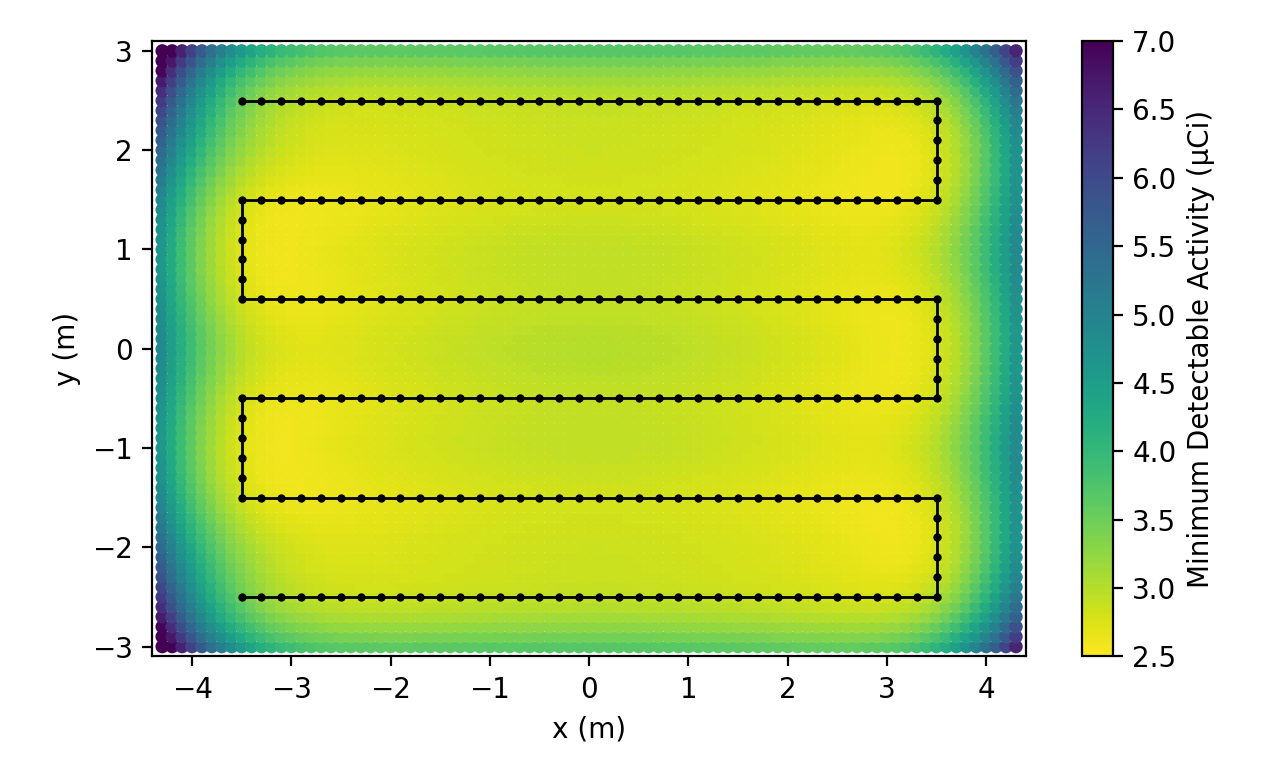}
\end{center}
\caption{The map of MDA values found for the toy problem.\label{fig:toy_mda}}
\end{figure}

The toy problem provides an opportunity to test whether the overall method results in the desired false positive rate (5\%) under the true statistics of the problem and not a simplification.
To do this test, 2,000 random resamples of the measurements were performed and PSL was applied to each one for all test points, and the PSL-derived source activities were compared to the critical values.
Using the single-point value for \(k_{\alpha}\), 1.645, the false positive rate of the samples was 75.1\%.
Using the corrected \(k_{\alpha}\) value, 3.023, the false positive rate was 7.6\%, much closer to the goal of 5\%.
An asymptotic correction factor of \(\eta = 0.05\) brought the false positive rate to 5.2\%, which was statistically consistent with the targeted 5\%.
Using these final values for \(k_{\alpha}\) and \(\eta\), the MDA was calculated and is shown in~\Fref{fig:toy_mda}.

\begin{figure}
\begin{center}
    \includegraphics[width=0.99\columnwidth]{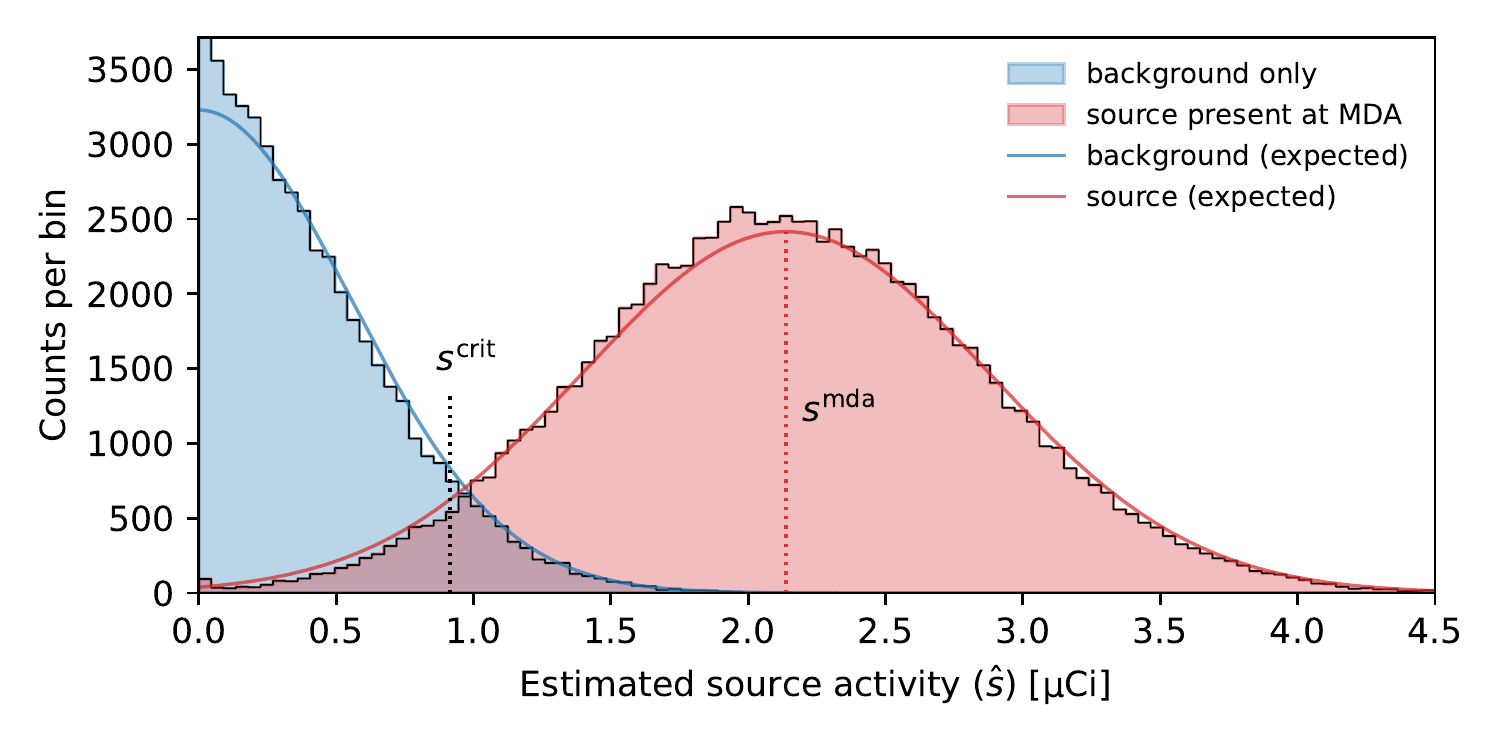}\\
    \includegraphics[width=0.99\columnwidth]{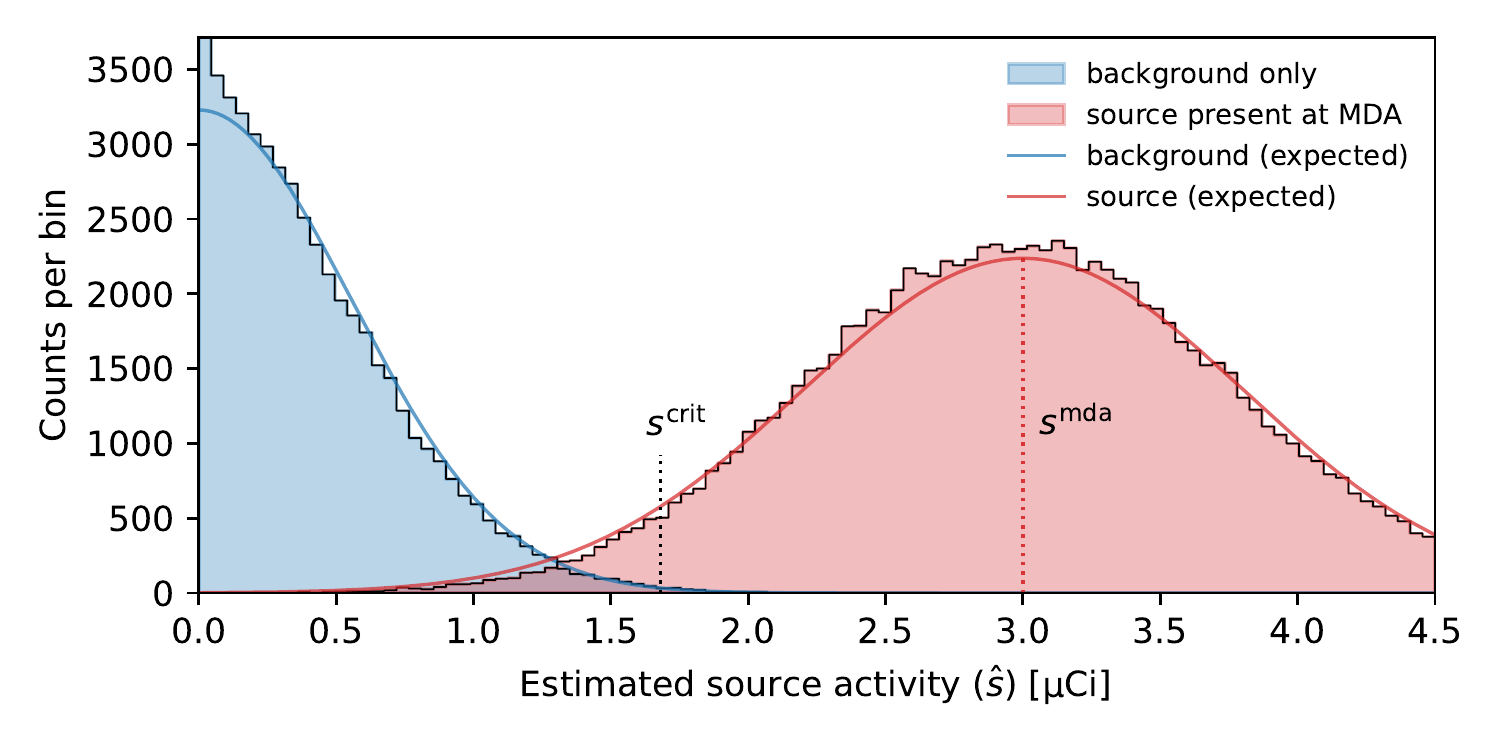}
\end{center}
\caption{Histograms of the estimated source activities for a test point at the origin in the toy problem when the data are truly from a constant background (blue) and when the data are from a constant background with a point source at the origin and with an activity equal to the MDA implied by the background rate (red).
The top histograms assume a single test point when calculating the critical value, while the bottom histograms increase the critical value to account for the presence of all of the other test points.
The expected distributions are Gaussian with means of \(0\) and \(s^{\mathrm{mda}}\) and standard deviations of \(\sigma_{\hat{s}}(b, 0)\) and \(\sigma_{\hat{s}}(b, s^{\mathrm{mda}})\), respectively, with some portion of the distributions ``piling up'' at zero and resulting in delta functions due to the nonnegativity constraint on \(\hat{s}\).
An asymptotic correction factor of \(\eta = 0.05\) was used for all of the expected distributions. \label{fig:toy_hists}}
\end{figure}

As an additional consistency check, 2,000 random samples were also drawn where, in addition to background, there was a point source located at the test point at the origin with an activity equal to the calculated MDA for that point.
Two scenarios were considered --- the MDA for the single-point assumption, and the MDA including the multi-point correction.
The resulting source activity estimates were calculated and are plotted in~\Fref{fig:toy_hists} to show that the toy model is in general agreement with the theory sketched out in~\Fref{fig:mda_diagram}.
The only differences between the two scenarios is the value of \(k_{\alpha}\) used in the calculation of the critical value and the resulting value of \(s^{\mathrm{mda}}\).

\section{Experimental results}\label{sec:experimental-results}
The MDA estimation method described in~\Fref{sec:methods} was also applied to measured data.
The results that will be shown here use the NG-LAMP system, which consists of four 1\(\times\)1\(\times\)2-inch CLLBC detectors with a package of contextual sensors, including a camera, inertial measurement unit (IMU), and a light distancing and ranging (LiDAR) unit~\cite{pavlovsky_3d_2019}.
The system was hand-carried in three similar surveys through two rooms of a small building by an operator who had no knowledge of whether any sources were present nor their potential locations during the surveys.
The three  surveys were: (1) with no sources, (2) with one $\approx$8\,\textmu Ci \isot{Cs}{137} source on a desk, and (3) with two $\approx$8\,\textmu Ci \isot{Cs}{137} sources stacked in the same position as (2).
The IMU and LiDAR data were used to perform SLAM, which resulted in pose solutions (i.e., \(\mathbf{r}_i\) and \(\mathbf{q}_i\) at each measurement time) and 3-D point clouds.

\begin{figure}
\begin{center}
\includegraphics[width=0.99\columnwidth]{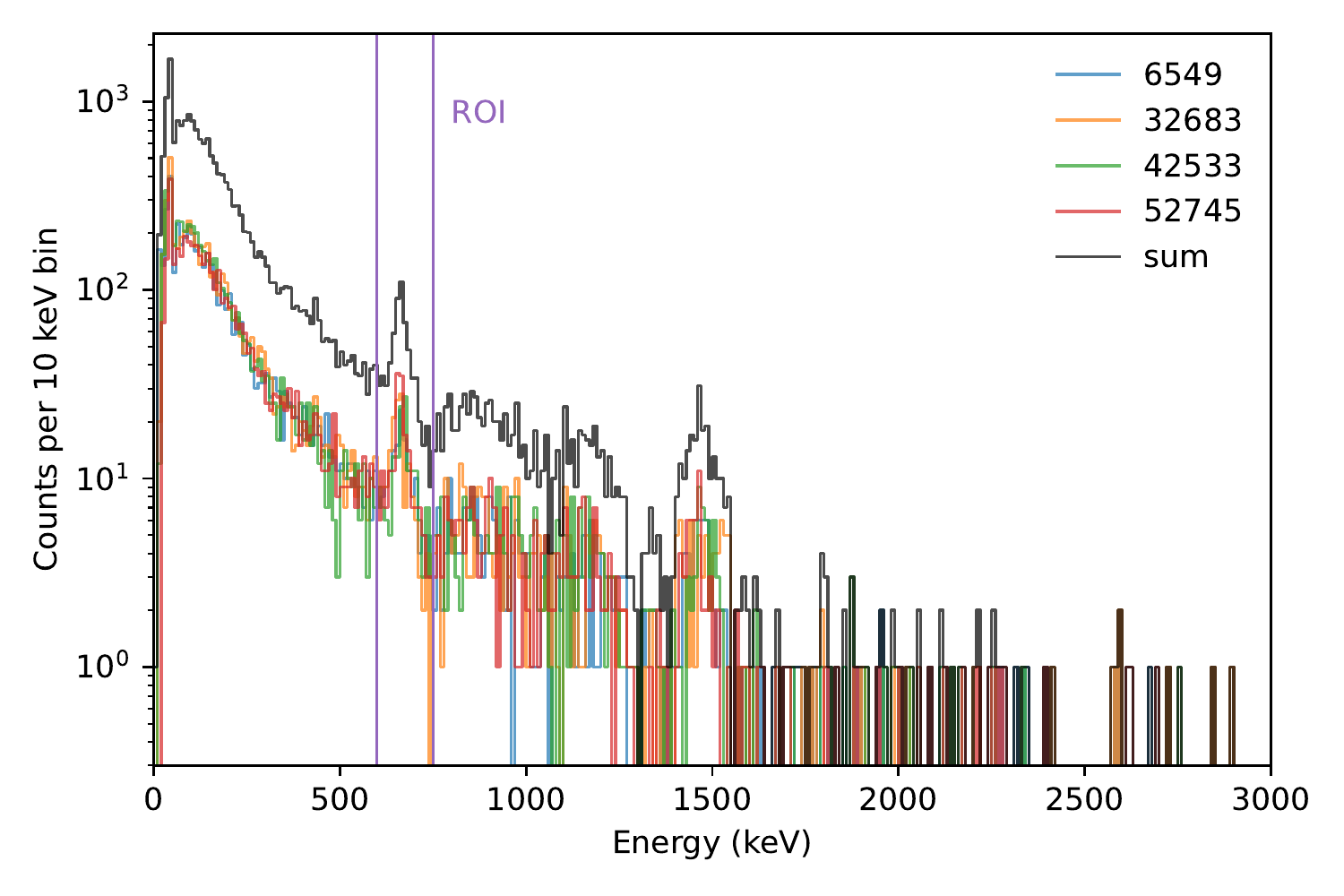}
\end{center}
\caption{The total spectrum for all four CLLBC detectors for run~3, which has the strongest source and lasted for 90\,s.
The spectral ROI was chosen to fully enclose the 662\,keV peak of \isot{Cs}{137}.
Also visible are other background features, such as the \(\approx\)30\,keV X-ray lines and 1436\,keV gamma-ray line of \isot{La}{138}, a primordial isotope present in the detector material.
This latter line forms a doublet with the 1460\,keV line from natural \isot{K}{40} decays.
\label{fig:spectrum}}
\end{figure}

Gamma-ray event data from the four CLLBC detectors were reported at a frequency of 4\,Hz, so \(\Delta t_i = 0.25\)\,s was used as the integration time for the method.
Event data were also kept separate by detector so that detector-specific responses could be used, enabling the exploitation of the shielding effects from neighboring detectors to provide crude directional information.
A spectral ROI of 600--750\,keV was chosen to analyze the data, a region wide enough to fully enclose the 662\,keV photopeak in all four detectors (\Fref{fig:spectrum}).
To calculate the response matrix, \( 4 \pi \) response functions derived from simulations and benchmarked with lab measurements were used to estimate the effective area of each detector~\cite{hellfeld_free-moving_2021}.
Test points were generated by voxelizing the point cloud at a pitch of 10\,cm and selecting the centers of any voxels that contained any LiDAR points (i.e., the ``occupied'' voxels).
In this way we obtained points that essentially covered the surface of the \mbox{3-D} office space.
For the three runs, between 61,000 and 68,000 test points were obtained in this way and used in the analysis.

\begin{figure}
\begin{center}
\includegraphics[width=0.99\columnwidth]{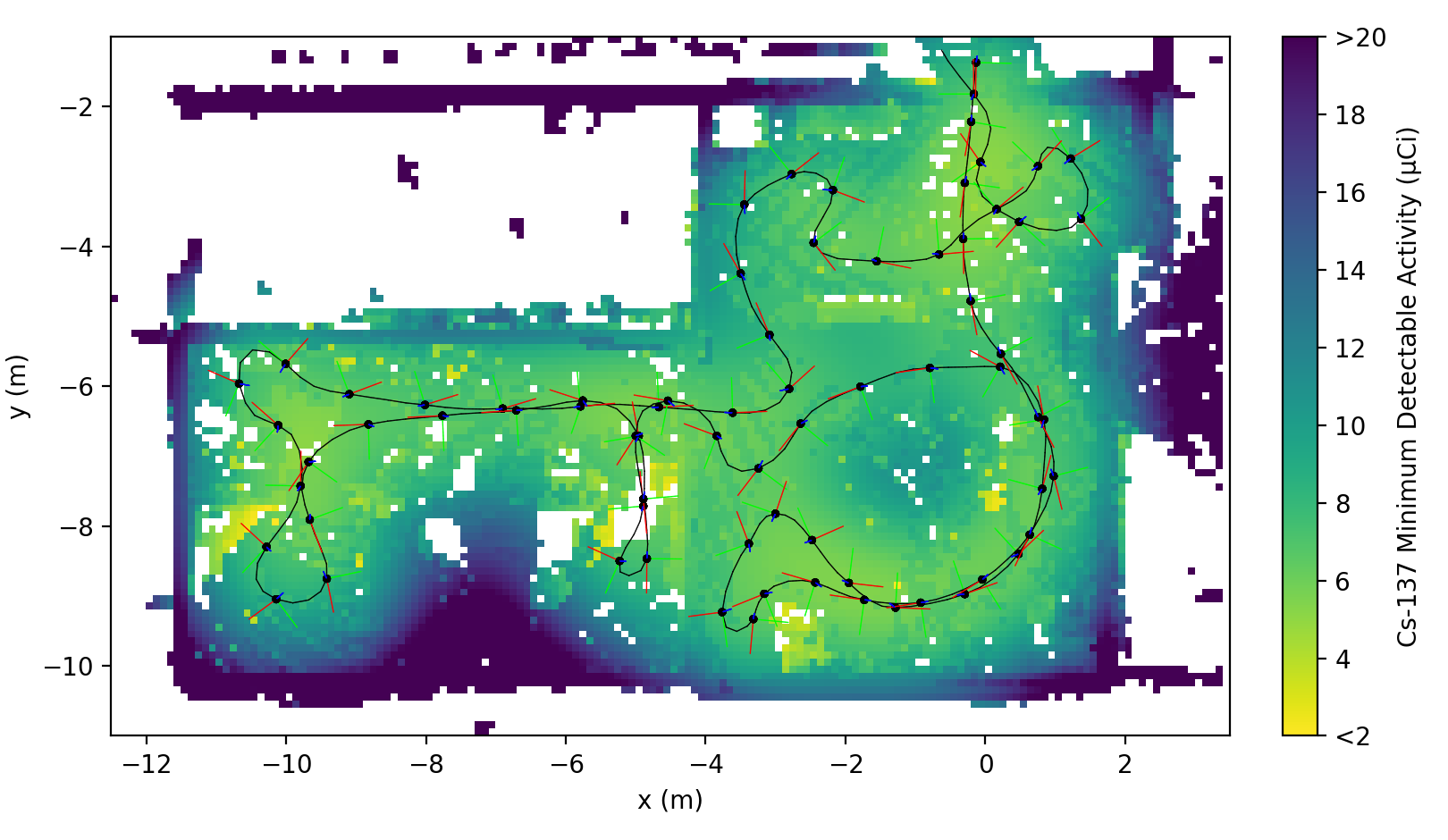}\\
\vspace*{0.55cm}
\includegraphics[width=0.99\columnwidth]{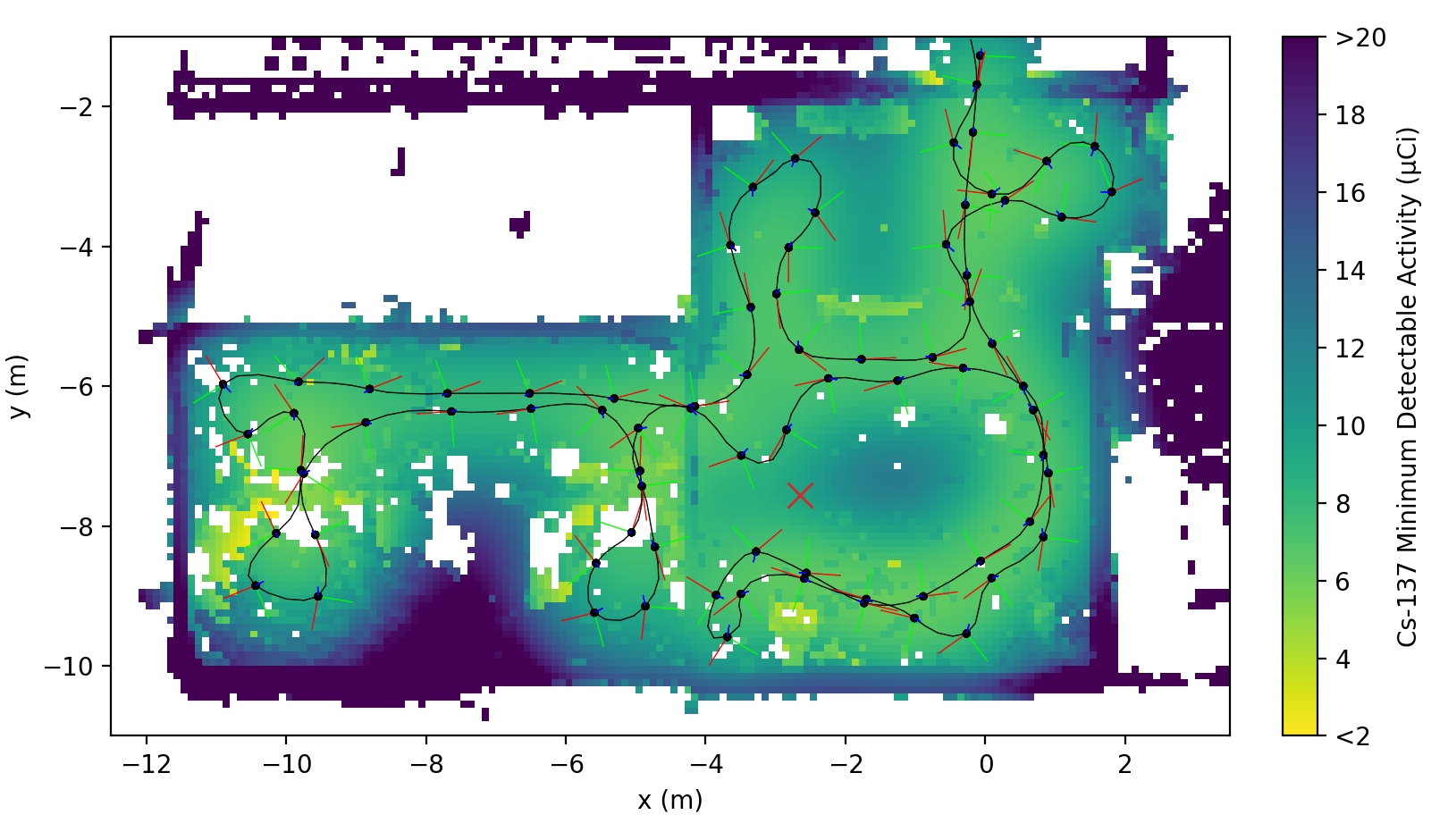}\\
\vspace*{0.55cm}
\includegraphics[width=0.99\columnwidth]{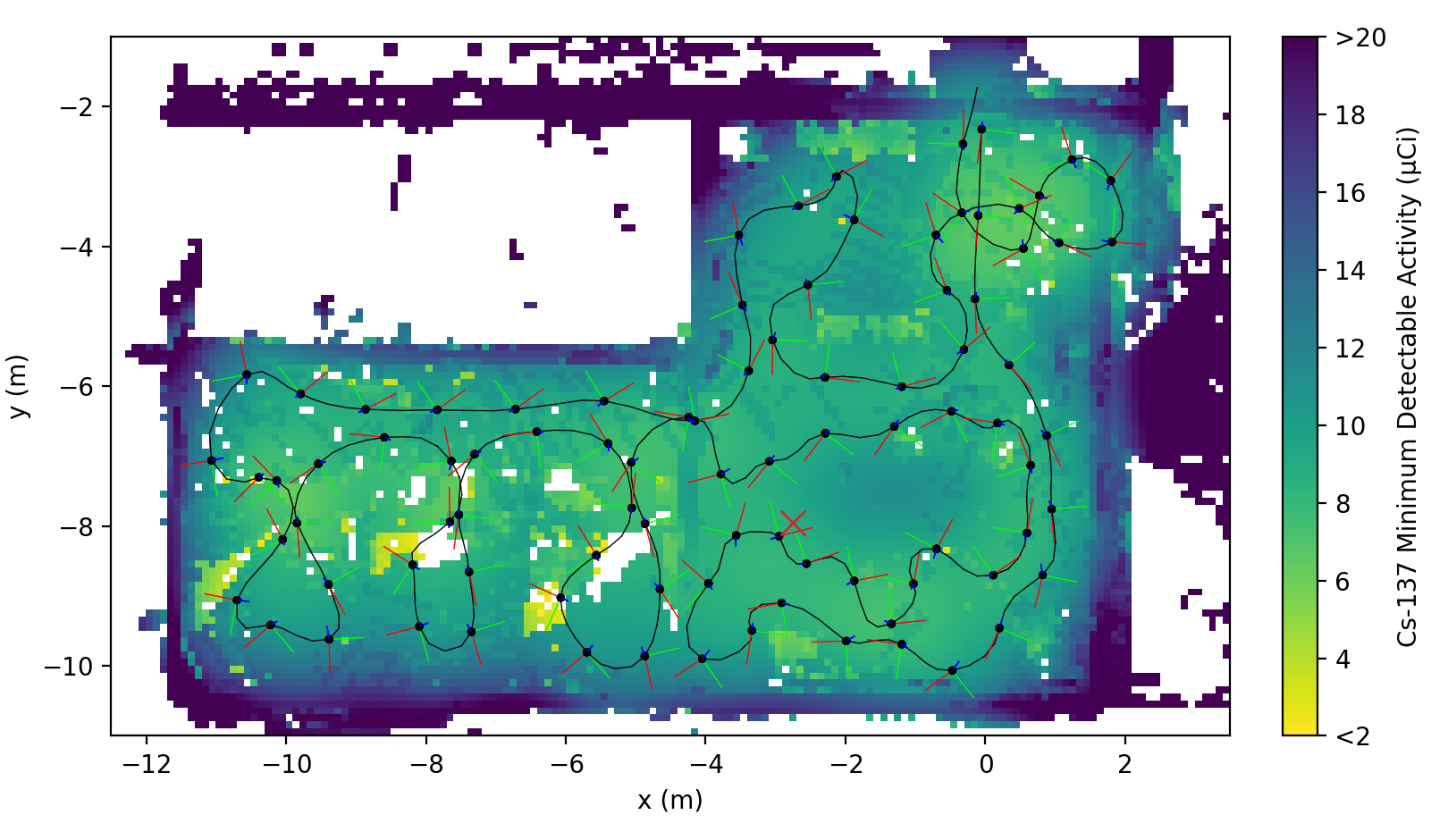}
\end{center}
\caption{Top-down view of the MDA maps calculated for each of the surveys, with the detector system's path and orientation annotated.
To obtain this 2-D projection from the 3-D map, the maximum MDA value along each vertical column is displayed.
White regions represent areas where there are insufficient LiDAR points to create occupied voxels anywhere in the vertical column.
For runs~2 and 3, where PSL solutions were found above the critical value, a red ``X'' marks the location of the maximum likelihood test point, which is consistent with the true location.\label{fig:mda_map}}
\end{figure}

The MDA approximation described in~\Fref{sec:methods} was applied to each dataset, and PSL was also calculated.
The resulting MDA maps are shown in~\Fref{fig:mda_map}.
Because sources were present for runs~2 and 3, the mean count rate \(\bar{b}\) was higher than run~1, and the overall MDA levels were accordingly slightly higher.
For all three runs, \(J\)~=~1,000 random samples were used to estimate the corrected \(k_{\alpha}\), and \(K\)~=~26, 29, and 31 eigenmodes were needed to explain 99\% of the eigenmode variances for the three runs, respectively.
The resulting \(k_{\alpha}\) values were approximately 3.0 (cf. the single-point value of 1.645).
An asymptotic correction factor of \(\eta = 0.05\) was used since the minimum number of detected events (282 for run~1) was similar to the expected number of events in the toy model (236).

\begin{figure}
\begin{center}
\includegraphics[width=0.99\columnwidth]{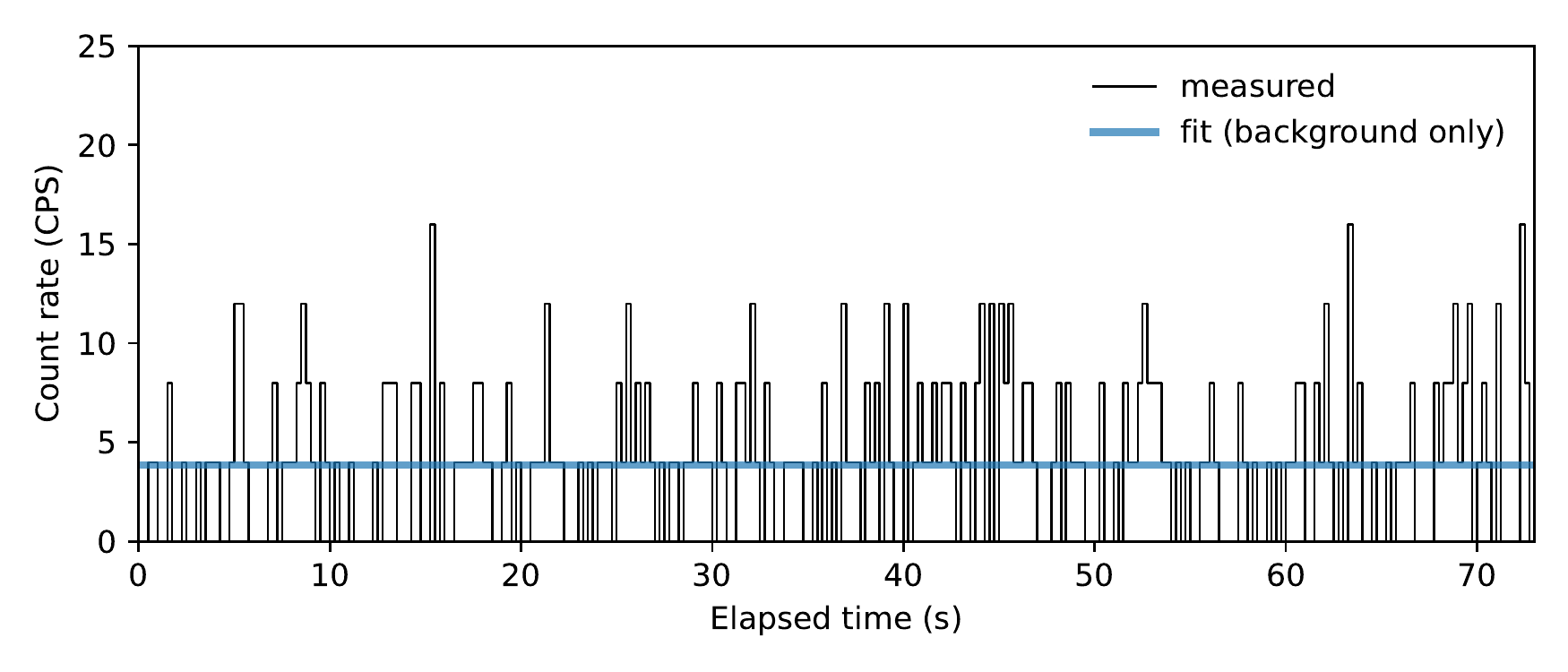}\\
\includegraphics[width=0.99\columnwidth]{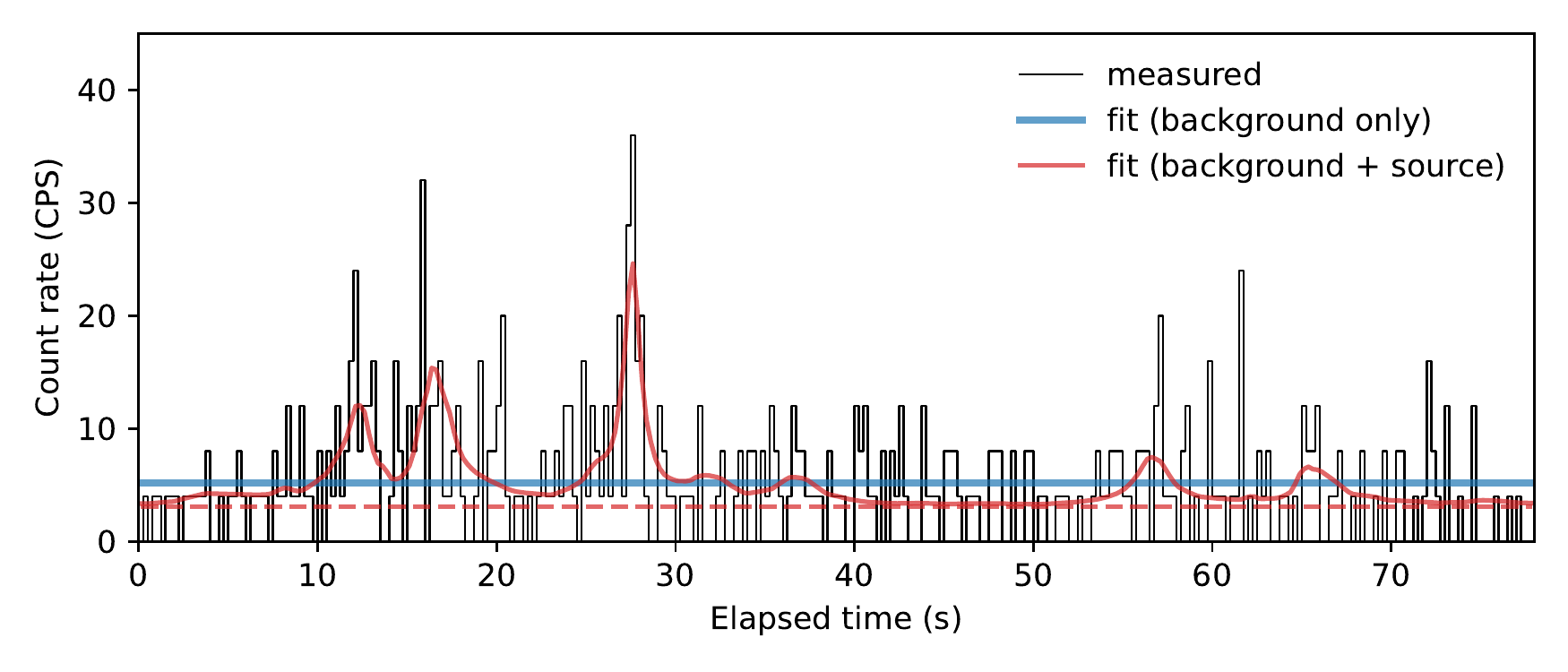}\\
\includegraphics[width=0.99\columnwidth]{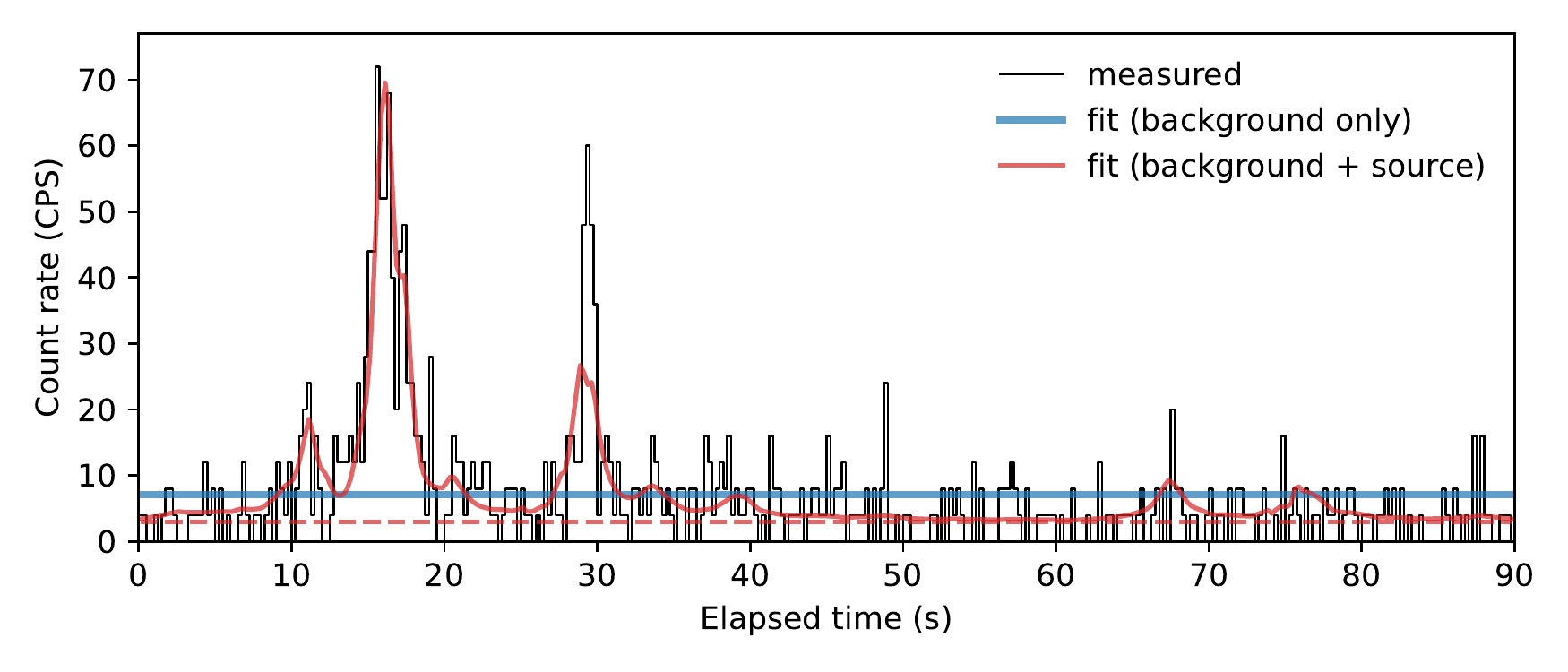}
\end{center}
\caption{The measured count rates in the spectral ROI for each of the three runs, with fits assuming background only and background with the most likely source.
For simplicity the count rates shown are summed over all four detectors, but the analysis has been performed without the detector measurements co-added.
Execution of the MDA algorithm on run~1 (top) resulted in no source found above the critical value, so no source fit is shown. \label{fig:counts_fit}}
\end{figure}

\begin{figure*}
\begin{center}
\includegraphics[width=0.75\textwidth]{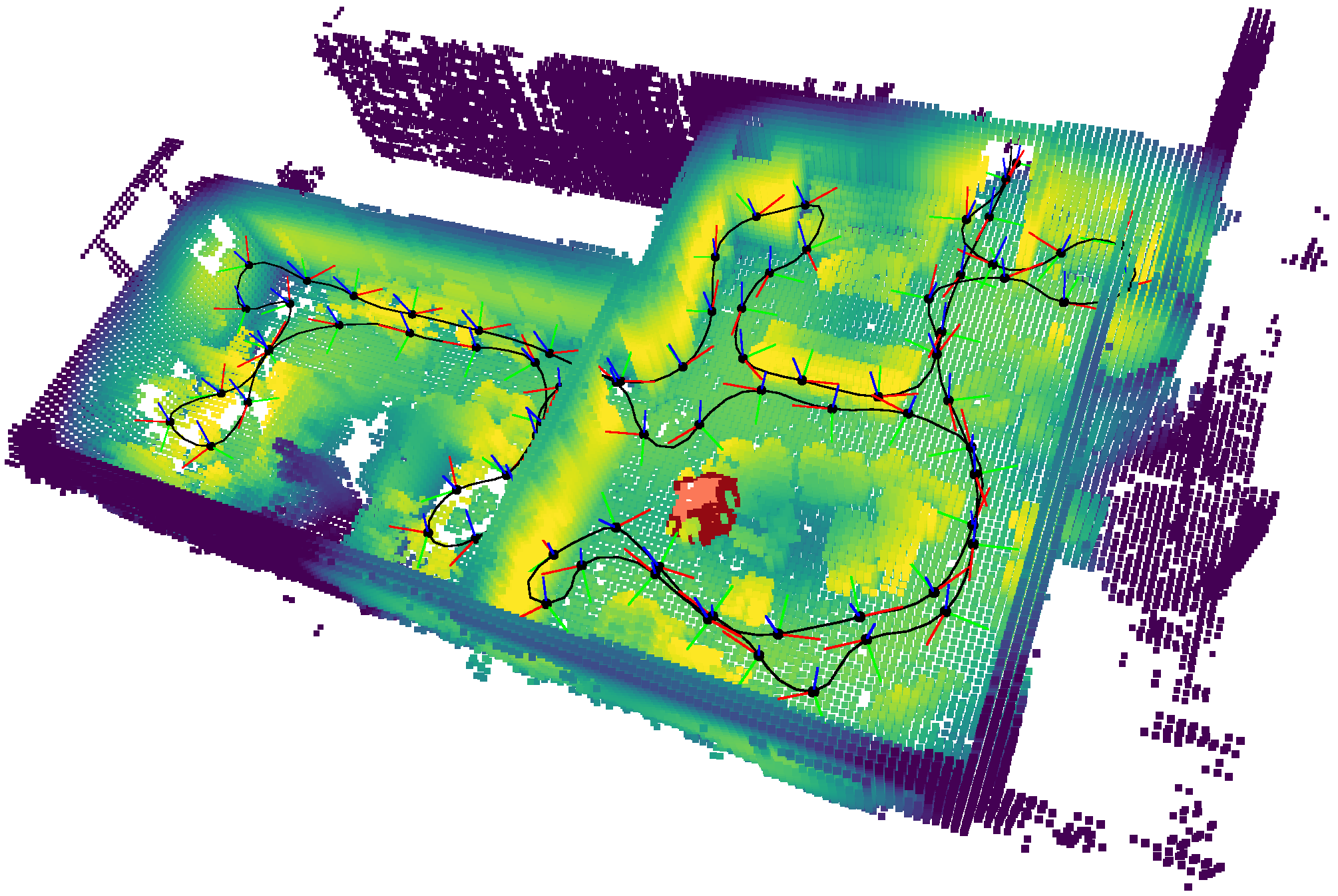}
\includegraphics[width=0.10\textwidth]{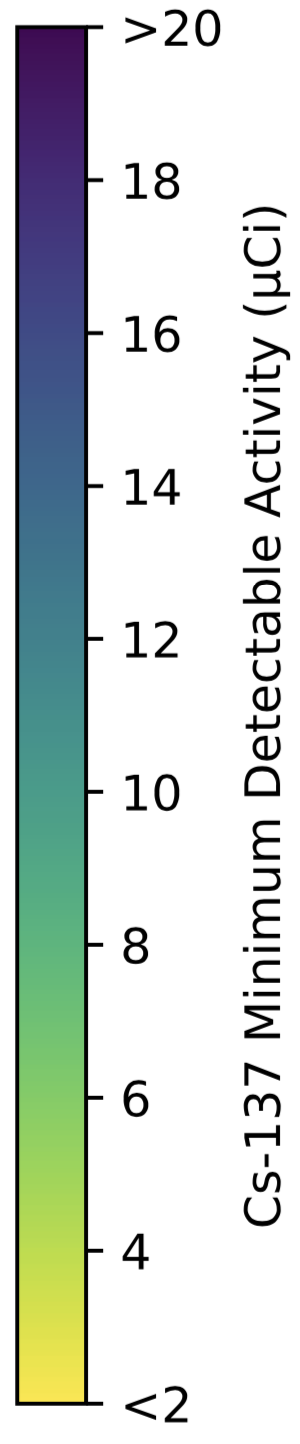}
\end{center}
\caption{Three-dimensional view of the MDA map for run~2, shown with the system pose solution (black lines and points show the position of the track, and red, green, and blue axes denote the orientation) and the PSL source localization 95\% confidence interval (shades of red and orange).\label{fig:3d_view}}
\end{figure*}

For run~1, no test point had a PSL-derived source activity above the calculated critical value, which was consistent with the ground truth (no source present).
For both runs~2 and 3, point sources were (correctly) found above the corresponding thresholds, and the 95\% spatial confidence intervals included the true locations and 95\% activity confidence intervals included the true activities.
\Fref{tab:results} summarizes the results of these analyses, \Fref{fig:counts_fit} shows the best fits to the measured count rates, and \Fref{fig:3d_view} shows a \mbox{3-D} view of the MDA map produced for run~2 with the source location confidence interval from PSL overlaid.
Although this test is limited (only one trial of each of three scenarios), it does demonstrate the feasibility and potential usefulness for the method in field measurements.

\begin{table*}
\caption{PSL results for the three measurements.\label{tab:results}}
\centering
\begin{tabularx}{0.95\textwidth}{XXXXX>{\raggedright}X>{\raggedright}XX}
    Run name & Duration [s] & Total counts  & \(\bar{b}\) [cps] & Corrected \(k_{\alpha}\) & True source activity [\textmu Ci] & \(\hat{s}_{j_{\mathrm{max}}}\) [\textmu Ci] & 95\% Confidence interval [\textmu Ci] \\
    \midrule
    run~1 & 73.0 & 282 & 0.97 & 2.99 & 0 & N/A & N/A \\
    run~2 & 78.0 & 404 & 1.29 & 3.04 & 7.7 & 11.2 & 4.8 -- 23.0 \\
    run~3 & 90.0 & 641 & 1.78 & 3.00 & 15.5 & 18.1 & 11.7 -- 27.1 \\
    \midrule
\end{tabularx}
\end{table*}

\section{Discussion}\label{sec:discuss}
With the advent of new technologies (handheld detectors with real-time SLAM capabilities), we have applied well known statistical principles (maximum likelihood and MDA theory) to a well studied problem (point source localization and quantification) to enable a new capability (the generation of MDA maps to describe measurement sensitivities of a \mbox{3-D} environment).
These maps could be generated in near-real time, leading to the potential of using such maps during search operations so that an area can be ``cleared'' by, e.g., ensuring the MDA at every point in the area is below a chosen MDA value.
This capability will enhance the situational awareness of operators performing search in complex environments.
It could also be used to direct autonomous survey instruments, where the system (e.g., a drone-mounted detector) could be instructed to choose pathways through the scene until the entire area has been cleared to a chosen MDA\@.

At the same time, we have necessarily had to elide some real-world complexities to make the method practical, and these issues should be carefully considered when applying this work.
We have relied on a single spectral ROI capturing an isotope's photopeak, and the related assumption that the background in the spectral ROI be constant.
This assumption may often be valid for small detectors in areas where the background does not change much, but there can be situations where the background varies too much (here PSL will also fail).
To improve the robustness of this approach (and also PSL), one could consider analyzing more parts of the spectrum than a single ROI.
For example, one or more additional spectral ROIs could be used to estimate the background within the primary ROI, or perhaps a technique could be developed to use the entire spectrum outside the ROI to estimate the background within the ROI.

Additionally, computational speed was not optimized here but would have to be for a real application.
For run~3, which was 90\,s long and used \(\approx\)68,000 test points, a CPU implementation of PSL alone took 41\,s, and the MDA estimation took an additional 250\,s, also on a single CPU\@.
Most of the MDA calculation time was spent performing SVD (148\,s) to obtain \(M=1440\) eigenmodes, while only \(K=31\) were needed to explain 99\% of the variance.
Significant speedup could be obtained by an iterative approach to SVD, stopping once the explained variance is high enough, since the total variance can be known beforehand (it is the trace of \(\var[\hat{\mathbf{s}}]\)).
In this case, performing SVD for only the first \(K\) eigenvectors was ten times faster (15\,s).
The second largest step by time usage (40\,s) was the iterative solution to~\fref{eq:s_mda}.
A simple bisection search was implemented that took 70\,s to reach an MDA accuracy of 1\% of the critical value.
The calculation time was dominated by the repeated calculation of \(\sigma_{\hat{s}_j}(b, s)\) for each of the test points.
This step would benefit from another method that more rapidly converges to the result, perhaps an implementation of Newton's method using numerical derivatives, and more accurate initial estimates.

Care must also be taken in the non-asymptotic approximation used here, captured by the factor \(\eta\).
In most practical applications, one cannot accrue enough statistics to assume the standard deviations of the estimated parameters are exactly equal to the CRLB\@.
To quantify the exact departure from asymptotic behavior, one could perform bootstrapping, a time-consuming process~\cite{wan_detection_2012}.
To develop a methodology that could be performed in real time or near-real time, the \(\eta\) factor was introduced in~\fref{eq:sigma-s-hat} to allow the method to be sufficiently accurate.
We expect that appropriate values for \(\eta\) for a given system and measurement can be estimated using representative toy problems, as was done here.

A final additional complexity that has been elided is that the MDA values presented here are calculated for unshielded, bare sources at each test point, but the influence of attenuation and scattering by passive material in the environment can change the spectral signatures seen by the detector.
If assuming no attenuation, estimated MDA values may end up lower than they would be if attenuation were taken into account, which could mislead an operator into thinking an area has been cleared that has not been.
There has been some work to include attenuation within PSL~\cite{bandstra_improved_2021}, but more research is needed to understand how to provide a useful range of MDA estimates given a \mbox{3-D} scene.

A potential extension of this work lies in adapting the method to calculate the MDA in the scene \textit{after} a source has been detected.
This situation is highly relevant to operations --- if an area were being surveyed and a source were found somewhere in it, the current method does not allow one to set maximally stringent limits on the rest of the scene, owing to the incorporation of source counts into the background estimate.
One way to extend the method would be to fix the maximum likelihood test point index \(j_{\mathrm{max}}\) and activity (\(\hat{s}_{j_{\mathrm{max}}}\)).
This assumption is sensible since once the first source is found, the operator (or autonomous system) would be able to focus on measuring the area immediately around that source, thus giving high confidence in the location and activity of the source.
The same analysis steps could be followed as presented in~\Fref{sec:methods}, but now incorporating the first point source contribution into the count rate estimate \(\lambda_{ij}\), and continuing to fit new values of \(b\) and \(s\).
Then \(s\) would become the activity of a potential second source in the scene.
Various other extensions of this MDA method that exploit the latest 3-D contextual information available from freely moving systems may also be possible.


\bibliographystyle{IEEEtran}
\bibliography{mda_mapping}


\end{document}